\documentclass[preprint,12pt]{elsarticle}

\usepackage{multirow}
\usepackage{booktabs}
\usepackage{longtable}
\usepackage{array}
\usepackage{adjustbox}
\usepackage{tabularx}
\usepackage{makecell}
\usepackage{listings}
\usepackage{nicematrix}
\usepackage{colortbl}
\usepackage{tcolorbox}
\usepackage{xcolor}
\usepackage{hyperref}

\newcommand{\sysname}{\textsc{Feed-O-Meter}}
\newcommand{\mentee}{\textit{Alex}}

\newcommand{\ipstart}[1]{\vspace{1mm}}

\definecolor{questioncolor}{HTML}{DEEBA6}
\definecolor{statementcolor}{HTML}{F9D9E6}

\definecolor{revision}{HTML}{BF1656} 

\tcbset{
  boxparamquestion/.style={
    colback=questioncolor,
    colframe=questioncolor,
    boxrule=0mm,
    arc=1mm,
    left=1mm,
    right=1mm,
    top=0.5mm,
    bottom=0.5mm,
    boxsep=0mm,
    valign=center
  }
}

\tcbset{
  boxparamstatement/.style={
    colback=statementcolor,
    colframe=statementcolor,
    boxrule=0mm,
    arc=1mm,
    left=1mm, 
    right=1mm,
    top=0.5mm,
    bottom=0.5mm,  
    boxsep=0mm,   
    valign=center 
  }
}

\newcommand{\question}[1]{%
  \tcbox[boxparamquestion, on line]{\textsf{#1}}%
}

\newcommand{\stat}[1]{%
  \tcbox[boxparamstatement, on line]{\textsf{#1}}%
}

\journal{International Journal of Human-Computer Studies}

\begin{document}

\begin{frontmatter}

\title{\sysname{}: Investigating AI-Generated Mentee Personas as Interactive Agents for Scaffolding Design Feedback Practice}

\author[inst1]{Hyunseung Lim}
\ead{charlie9807@kaist.ac.kr}
\author[inst1]{Dasom Choi}
\author[inst2]{DaEun Choi}
\author[inst1]{Sooyohn Nam}
\author[inst1]{Hwajung Hong\corref{cor1}}
\ead[URL]{https://dxd-lab.github.io/}

\affiliation[inst1]{organization={KAIST, Department of Industrial Design},
            city={Daejeon},
            country={Republic of Korea}}

\affiliation[inst2]{organization={KAIST, School of Computing},
            city={Daejeon},
            country={Republic of Korea}}
            
\cortext[cor1]{Corresponding author.}

\begin{abstract}
Effective feedback, including critique and evaluation, helps designers develop design concepts and refine their ideas, supporting informed decision-making throughout the iterative design process. However, in studio-based design courses, students often struggle to provide feedback due to a lack of confidence and fear of being judged, which limits their ability to develop essential feedback-giving skills. Recent advances in large language models (LLMs) suggest that role-playing with AI agents can let learners engage in multi-turn feedback without the anxiety of external judgment or the time constraints of real-world settings. Yet prior studies have raised concerns that LLMs struggle to behave like real people in role-play scenarios, diminishing the educational benefits of these interactions. Therefore, designing AI-based agents that effectively support learners in practicing and developing intellectual reasoning skills requires more than merely assigning the target persona's personality and role to the agent. By addressing these issues, we present \sysname{}, a novel system that employs carefully designed LLM-based agents to create an environment in which students can practice giving design feedback. The system enables users to role-play as mentors, providing feedback to an AI mentee and allowing them to reflect on how that feedback impacts the AI mentee's idea development process. A user study (N=24) indicated that \sysname{} increased participants' engagement and motivation through role-switching and helped them adjust feedback to be more comprehensible for an AI mentee. Based on these findings, we discuss future directions for designing systems to foster feedback skills in design education.
\end{abstract}



\begin{keyword}
Design Education \sep Design Feedback \sep Human-Computer Interaction \sep Large Language Model \sep AI-Generated Agent
\end{keyword}

\end{frontmatter}

\section{Introduction}
In the iterative design process, where solutions are progressively refined, feedback is indispensable for enhancing the quality and effectiveness of design~\cite{wynn2017perspectives}. Mastering the art of giving feedback is critical for designers, as it not only improves design performance~\cite{wynn2022feedback} but also integrates the diverse perspectives within a design team towards a shared objective~\cite{valkenburg1998reflective} while enabling the expression of individual design viewpoints~\cite{braha1998measurement}. Recognizing its significance, design education has long emphasized feedback as a core learning experience. Among various educational strategies, peer feedback is a common way for students to practice giving and receiving feedback, which raises the quality of their work while fostering key competencies such as communication, collaboration, and critical thinking~\cite{mcdonnell2016scaffolding, bjorklund2004effects}. 

Despite the recognized value of feedback in enhancing both design outcomes and student competencies, there remain few opportunities for students to learn how to provide effective feedback. While many design studio courses encourage peer feedback through project-based learning~\cite{ching2013peer, ertmer2007using}, students often face challenges in providing constructive feedback and remain disengaged in peer feedback~\cite{ching2013peer, gielen2010improving, hovardas2014peer}. This difficulty stems from students' limited feedback experience and design knowledge~\cite{ching2013peer}, as well as anxiety or hesitation about giving feedback due to concerns over receiving criticism for inadequate input~\cite{ertmer2007using, gielen2010improving, cook2020designing}. To address this challenge, previous studies have emphasized the importance of providing comfortable educational environments that encourage students to actively engage in giving feedback and cultivate their feedback skills~\cite{cook2020designing, jug2019giving}.

The HCI community has sought to create environments that foster higher-quality feedback in creative activities such as design. Prior research has proposed online platforms~\cite{cheng2020critique, lambropoulos2013hci} that allow design students to exchange feedback at scale, particularly crowd-sourcing systems~\cite{oppenlaender2021hardhats, krause2017critique,lekschas2021ask}, and interactive guidelines for crafting effective feedback~\cite{ngoon2018interactive}. Although these approaches increase the volume of feedback and participation~\cite{cheng2020critique, oppenlaender2021hardhats}, limited motivation and the absence of actual bi-directional communication often lead to superficial feedback~\cite{oppenlaender2021hardhats, nguyen2017fruitful}. Meanwhile, advances in large language models (LLMs) have prompted investigations into AI agents that help learners practice logical and critical thinking skills—including arguing~\cite{wambsganss2021arguetutor}, teaching~\cite{markel2023gpteach}, speech practice~\cite{park2023audilens}, and conversing~\cite{shaikh2024rehearsal}. For instance, GPTeach~\cite{markel2023gpteach} employs an LLM-driven agent that adopts a student persona, enabling teaching assistants to rehearse their instructional strategies through simulated teaching scenarios. Role-playing with AI agents lets learners engage in multi-turn feedback without the anxiety of external judgment or the time constraints of real-world settings~\cite{wambsganss2021arguetutor, markel2023gpteach, shaikh2024rehearsal}.

However, prior studies have raised concerns that LLMs struggle to behave like real people in role-play scenarios, thereby diminishing the educational benefits of these interactions~\cite{jin2024teach, lim2024co, jo2023understanding}. For example, LLM-based agents with a student persona often provide responses that are far more intelligent and refined than those of actual students due to the extensive knowledge embedded in the LLMs~\cite{jin2024teach, lim2024co}. Studies revealed that this misalignment between the agent's capabilities and its intended persona can lead users to inadvertently rely on the AI's responses rather than developing their own abilities through interactions~\cite{jin2024teach, lim2024co, yen2024give}. Therefore, designing AI-based agents that effectively support learners to practice and develop intellectual reasoning skills requires more than merely assigning the target persona's personality and roles to the agent; it demands deliberate strategies to align the AI's behavior and knowledge system with its intended educational role~\cite{markel2023gpteach, jin2024teach, jo2023understanding}.

Building on these insights, this study introduces a novel system with AI agents that empowers students to practice their design feedback skills. We propose \sysname{}, a system that facilitates design feedback practice through role-playing interactions, allowing students to mentor an AI that adopts the persona of a design student. The system includes key features: (1) a chat interface for role-switched conversations, users act as mentors and the AI as a mentee, (2) feedback reflection interfaces that show how their feedback influences the AI mentee's idea development, allowing users to reflect on the effectiveness of their feedback and make adjustments accordingly. By incorporating LLMs, \sysname{} enables the AI agent to understand and respond to feedback in real-time while maintaining the role of a design student and visually demonstrating how the feedback impacts the mentee's idea development process. 

To examine how students interact with \sysname{} and its effectiveness in enhancing design feedback skills, we conducted a user study with 24 design students. This study employed a within-subject comparative study to assess the impact of the system's feedback reflection features on students' feedback. Our findings reveal that the AI mentee and \sysname{} environment provided a realistic and low-pressure feedback experience, allowing participants to engage more actively in feedback activities without fear of judgment. Moreover, the system's feedback reflection features encouraged participants to focus not only on the content of their feedback but also on effective communication strategies to make their feedback clear and acceptable. Participants recognized that \sysname{} could go beyond improving design feedback skills to help them critically analyze their own designs and identify areas for improvement. Based on our findings, we discuss the implications of role-playing interactions for feedback practice and offer insights into designing AI personas that effectively integrate these interactions within educational contexts.

The contributions of our paper are as follows: 
\begin{enumerate}
    \item The design and development of \sysname{}, a system that allows design students to improve their feedback skills through role-playing interactions with an AI mentee. We outline the system's design rationale and capabilities, demonstrating how it facilitates effective feedback practices.
    \item An empirical understanding of how design students engage with \sysname{}. Through a user study with 24 design students, we analyzed interaction logs and interview transcripts to provide insights into how students use the system to practice and refine feedback on design ideas.
    \item Design considerations for incorporating LLM-driven systems in design education. We discuss the benefits of leveraging LLMs to create interactive environments and AI agents that support learning design principles, as well as the challenges of simulating realistic design feedback scenarios.
\end{enumerate}

\section{Related Work}
\subsection{Interactive Tools for Design Education}
The HCI community has long explored the use of technology to support and enhance the design process~\cite{frich2021digital}. A key focus has been on developing creative support tools~\cite{shneiderman2002creativity} that promote effective engagement in inherently creative tasks, with design being a prime example. These studies have extended to educational technologies aimed at enhancing practical design skills. Several studies have introduced physical interactive tools for gaining specialized knowledge through practical experiences, such as haptic interfaces~\cite{minamizawa2012techtile}, IoT-based systems~\cite{jang2018development}, and physical computing platform~\cite{bianchi2024blinkboard}. Interactive applications in visual and graphic design education focus on enhancing learning through visual examples. For instance, learner-centered online design galleries allow students to acquire knowledge from curated examples~\cite{yen2022seeking}, while ProcessGallery~\cite{yen2024processgallery} helps users compare pairs of design examples to grasp key principles. DesignQuizzer~\cite{peng2024designquizzer}, an AI agent, assists users in gaining visual design knowledge by drawing on insights from an online community. 

While acquiring and applying design knowledge is valuable, many researchers in design education also emphasize that strengthening students' critical thinking serves as the backbone of effective and self-directed design practice~\cite{henriksen2017design, razzouk2012design}. This competency can be developed when designers engage in interactive feedback exchanges with stakeholders, continuously reflecting on the feedback received and applying it to refine their own ideas~\cite{zhu2014reviewing, ahern2019literature}. Yet, despite widespread discussions about the potential of digital tools and computer-supported creativity in design education, relatively little attention has been devoted to fostering the design feedback skills essential for nurturing critical thinking. Roldan et al.~\cite{roldan2020opportunities} noted that most design tools, methods, and guidelines used in design and HCI education concentrate on the act of designing while overlooking the reflection, such as feedback skills needed to assess outcomes. Given that reflective practice, such as peer feedback, has already been extensively explored in pedagogy for critical thinking, they also proposed integrating those strategies into design education tools~\cite{roldan2020opportunities, clemente2016learning}.

To bridge this gap, our study proposes educational tools that cultivate critical thinking skills in design, focusing on helping students practice and deliver effective feedback. In Section \ref{sec:feedback_design}, we review prior HCI research focused on eliciting high-quality design feedback. These studies focus on helping designers receive high-quality feedback, but there are still limitations in the educational aspect of fostering students' design feedback skills. In Section \ref{sec:AI_agents_for_learning}, we explore the potential of conversational agents in education and outline the requirements and design considerations for adapting these approaches to systems that foster students to provide better design feedback.

\subsection{Approaches to Improve Feedback in Design}
\label{sec:feedback_design}
Design feedback offers designers valuable insights and opportunities to refine their work~\cite{wynn2022feedback}. Nevertheless, obtaining high-quality feedback remains challenging because it requires feedback providers with deep design knowledge and extensive feedback experience, prerequisites for providing truly constructive and relevant critiques~\cite{ching2013peer}. Recognizing this challenge, researchers in HCI have proposed interactive tools and structured approaches that deliver effective feedback strategies and give feedback providers more practice applying them. One of the initial attempts to enhance the quality of feedback is to introduce structured guides and frameworks developed for effective design feedback~\cite{cook2020designing, krause2017critique, ngoon2018interactive, yuan2016almost}. Ngoon et al.~\cite{ngoon2018interactive} proposed CritiqueKit, an interactive guideline that provides rubrics (Specific, Justified, Actionable) with specific examples to enhance the quality of feedback. Krause et al.~\cite{krause2017critique} have further suggested that feedback guidelines generated by natural language models helped students better understand the characteristics of effective feedback and enabled them to provide more helpful feedback. One such approach introduced a scaffolding step that prompted students to reflect on their feedback before giving it, allowing them to provide more targeted and specific feedback~\cite{cook2020designing, greenberg2015critiki}. Recent research has even proposed strategies using LLMs to improve students' feedback by automatically adding positive summaries of feedback~\cite{yang2025understanding}. While these approaches have improved the quality of feedback, their effectiveness is limited in design education, where there are often no definitive answers due to the open-ended nature of design work. Manual feedback—such as rubric and structured guideline-based critiques—can unintentionally constrain the development of creative concepts by imposing rigid criteria~\cite{yuan2016almost}. Furthermore, creating guidelines and examples demands significant collaboration among experts and substantial effort to adapt them to different contexts~\cite{krause2017critique, yuan2016almost}. Many feedback strategies in previous studies take the form of static documents or written comments, rarely reflecting the dynamic nature of design feedback in conversational and practical settings.

In response to these constraints, research has shifted toward creating immersive environments that improve the quality of design feedback by encouraging active engagement~\cite{jug2019giving, lambropoulos2013hci, sadler1989formative}. Online communities have proven effective in enabling designers to provide design feedback by helping them overcome real-world constraints, such as anxiety or hesitation about giving feedback in a classroom setting~\cite{cheng2020critique, oppenlaender2021hardhats, krause2017critique, kang2018paragon}. For instance, Kang et al.~\cite{kang2018paragon} created Paragon, an online gallery that allows feedback providers to reference rubrics defined by recipients and tailor their comments accordingly, reducing social friction and ensuring the feedback addresses recipients' needs. While this approach has increased the frequency and involvement of providing feedback~\cite{cheng2020critique, oppenlaender2021hardhats}, concerns remain about the quality of feedback, as insufficient motivation and a lack of bi-directional communication often lead to superficial discussions~\cite{oppenlaender2021hardhats, nguyen2017fruitful}. Meanwhile, although fostering a competitive environment to prompt frequent feedback may boost student participation, it can also create discomfort by prompting students to compare themselves to one another and pressuring them to favor specific designs~\cite{cambre2018juxtapeer}. To overcome the inherent limitations of standard design feedback sessions, which require students to critique peers' work, this research aims to identify less burdensome environments that can provide more effective feedback for students and design novices.

\subsection{Conversational Agents for Education: From Automated Instruction to Persona-Enhanced Interactions} \label{sec:AI_agents_for_learning}
Advances in natural language processing (NLP) have paved the way for the AI-powered conversational agents that offer significant potential for educational support. While these agents have taken on multiple roles—from dictionary chatbots to automated graders~\cite{han2024llm}—the most compelling role is that of an AI tutor, which can automate instruction by simulating teacher-student role-play interactions~\cite{wambsganss2021arguetutor, han2024llm, graesser2004autotutor}. Prime examples include intelligent tutoring systems like AutoTutor~\cite{graesser2004autotutor}, which teach not only subjects ranging from physics to law but also basic learning capabilities such as reading comprehension~\cite{graesser2004autotutor, vanlehn2011relative, ma2014intelligent, kulik2016effectiveness}. Beyond these competencies, AI agents are advancing into advanced abilities like argumentation and critical thinking, exemplified by ArgueTutor, a dialogue-based system for teaching argumentation skills~\cite{wambsganss2021arguetutor}, CReBot for critical reading~\cite{peng2022crebot}, and Sara for video lecture comprehension~\cite{winkler2020sara}. Although these agents provide immersive experiences, previous research suggests that students risk becoming overly dependent on AI tutors, potentially hindering the development of essential higher-order thinking skills~\cite{fuchs2023exploring, yu2023reflection}. Over-reliance on AI tutors can prompt learners to forfeit critical evaluation of information quality, creative ideation, and the kind of critical thinking crucial for genuine intellectual growth~\cite{fuchs2023exploring}.

Recent breakthroughs in large language models (LLMs), such as ChatGPT, have further accelerated the evolution of these conversational agents, allowing them to mimic specific personas and create dynamic, context-aware conversations~\cite{han2023recipe, junprung2023exploring, baidoo2023education}. These studies have expanded their scope beyond conventional tutoring roles to address emerging concerns about student over-reliance and uncritical thinking. For instance, Algobo, an AI-powered teachable student, has been developed to teach programming skills using the learning-by-teaching theory~\cite{jin2024teach}. GPTeach also helps teaching assistants acquire teaching competencies by interacting with LLM-powered students~\cite{markel2023gpteach}. Other studies have explored role-playing interactions to enhance questioning skills, allowing users to pose more critical questions to AI students~\cite{lim2024identify}. Still, these approaches have encountered challenges, as the LLMs' extensive knowledge often led to overly adept responses that disrupted the student persona, diminishing the immersion in role-play and limiting educational effectiveness~\cite{jin2024teach, lim2024identify}. Such findings underscore the importance of not merely assigning specific personas to LLMs, but also carefully designing both the LLM pipeline and interactions to ensure that these personas function effectively~\cite {jin2024teach, lim2024identify}. To overcome these challenges, our research aims to explore how AI agents can be better implemented to support effective conversation with LLMs, including the use of a controlled knowledge state~\cite{jin2024teach}. Further details on the development of a system that facilitates the practice of design feedback through AI, focusing on key characteristics for effective design education, are provided in Section~\ref{sec:design_rationales}.

\section{Design of \sysname{}}

\begin{figure}[h]
  \includegraphics[width=\textwidth]{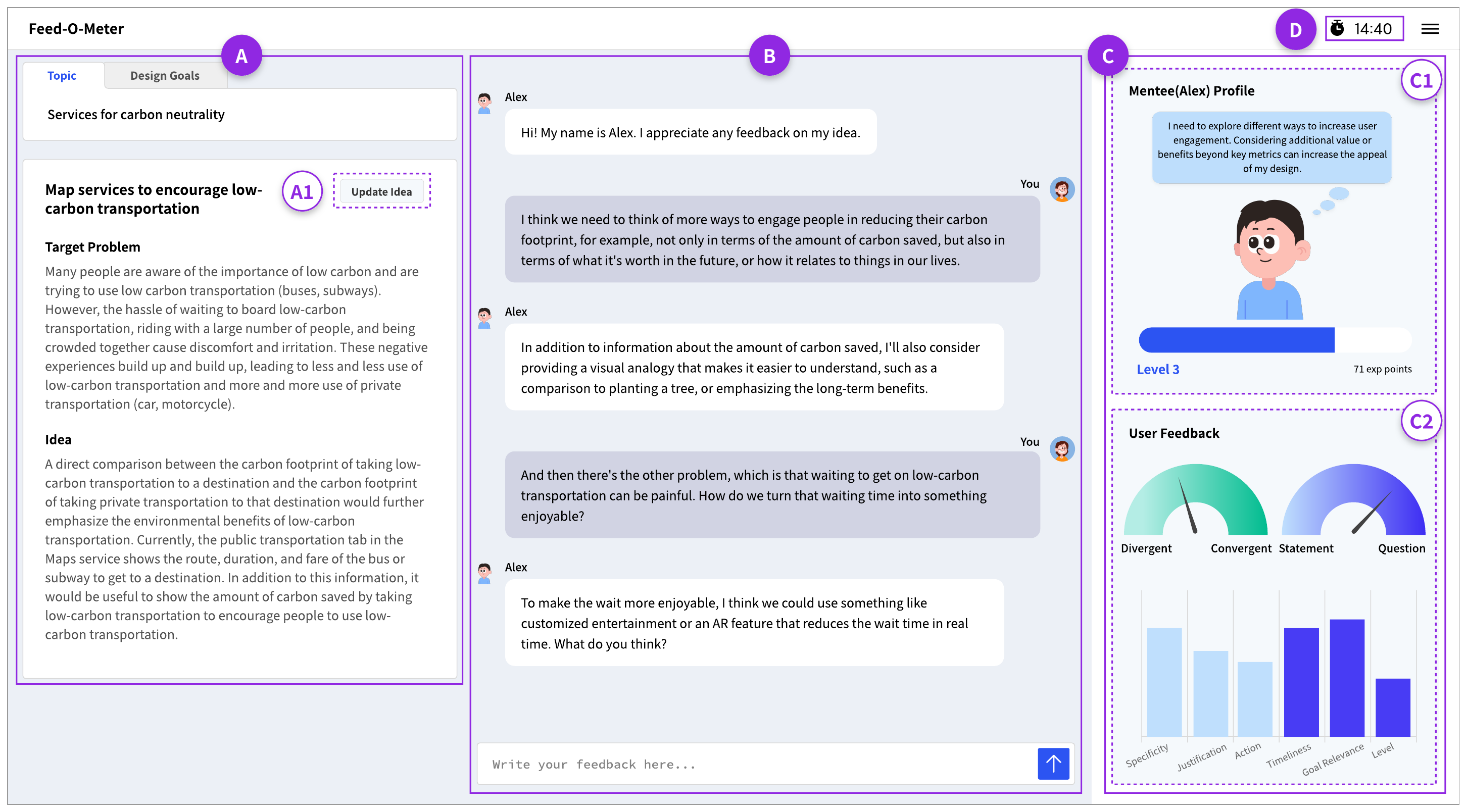}
  \caption[This image shows the Feed-O-Meter user interface, designed for role-playing feedback interactions with an artificial intelligence mentee. On the left, the Idea Proposal Interface displays the predefined design goals and the mentee's current design idea. In this case, the idea focuses on improving the waiting experience for low-carbon transportation. To the right of the idea is an "Update Idea" button, which updates the design based on received feedback and immediately reflects the changes in the interface. In the center, the Chat Interface allows users to provide feedback, with the mentee, Dave, explaining his ideas and requesting feedback on enhancing the transportation experience. On the right, the Intervention Interface is split into the Mentee Profile, showing Dave’s progress, and the Feedback Evaluation Dashboard, which visualizes feedback criteria such as timeliness, goal relevance, specificity, and justification. Finally, a timer at the top tracks the session duration.]{The main user interface of the \sysname{}. (A) The Idea Proposal Interface displays predefined design topics, goals, and the AI mentee's current design idea. By clicking the ``Update Idea'' button (A1), the user can request the AI mentee to update its design idea based on feedback. (B) Chat Interface allows users to provide feedback to the AI mentee. (C) Feedback Reflection Interface includes the Mentee's Profile (C1), showing the mentee's progress, and the Feedback Evaluation Dashboard (C2), which visualizes feedback criteria. The timer (D) tracks the session duration.}
  \label{fig:main_UI}
\end{figure}

We designed and developed a \sysname{}, which allows users to practice providing design feedback. This section provides a detailed description of the design rationales and the specifics of the system's pipeline.

\subsection{Design Rationales} \label{sec:design_rationales}
\subsubsection{DR1: Simulate a Novice Design Student as an Agent Persona}
As highlighted in related works, fostering students' feedback skills requires providing an immersive environment that encourages active engagement in the feedback process~\cite{jug2019giving, sadler1989formative}. Role-play has long been recognized as an effective teaching method, offering students indirect experiences of challenging situations~\cite{ahern2019literature}. By leveraging the capabilities of LLMs, which can simulate specific personas and facilitate role-playing interactions~\cite{junprung2023exploring}, we aimed to create a role-playing experience where users provide feedback during the design process.

Our objective was to enable users to practice providing feedback in scenarios that closely mirror real-life situations while fostering active engagement, in contrast to the hesitation often observed in traditional environments. To achieve this, we assigned users the role of a \textit{mentor} and designed scenarios where they provided feedback on an \textit{AI mentee's} design idea. This role-switching encourages users to adopt a new perspective and become more engaged in the task~\cite{ferrari2020sapeer, rao2012exploring} while also providing a learning experience similar to learning by teaching~\cite{fiorella2013relative}. In addition to simulating conversations where the AI mentee receives feedback, we also enabled behaviors that allow the mentee to update their ideas based on the feedback. Although students in the real world do not immediately revise their ideas, our system aims to practice feedback skills, enabling users to reflect on how effectively they have guided the mentee’s design ideas and prompting them to refine their own feedback strategies.

For these interactions to be genuinely effective, the AI mentee should be designed to gain knowledge and develop ideas based on user feedback rather than developing ideas on its own~\cite{jin2024teach, lim2024identify}. In line with this goal, we designed our AI Mentee to embody the persona of a novice design student with a knowledge state that advances exclusively through user-provided feedback. In our proposed scenarios, we distinguish two forms of knowledge state based on feedback type: \textit{knowledge} and \textit{action plan}. The \textit{knowledge} stores new information or evaluations drawn from user feedback, allowing the AI mentee to gradually build design expertise. The \textit{action plan} tracks recommendations for refining or updating the design ideas, enabling the AI mentee to update ideas based on user feedback.

\subsubsection{DR2: Promote Critical Reflections on Feedback and its Effects}
Our system is designed not only to provide an environment where users can practice giving feedback but also to help them improve their feedback skills. Rather than prescribing a specific feedback rubric, we emphasized user autonomy by allowing them to observe how their feedback influences the AI mentee’s design ideation process. We adopted the following three components to deliver indirect guidance and encourage users to reflect on the impact of their feedback and independently refine their strategies. 

First, the system provides real-time visual representations that assess both the type and quality of each piece of feedback. This design choice draws on evidence that nudging or visualizing message characteristics can encourage deeper self-reflection and improve overall feedback quality~\cite{shaikh2024rehearsal, menon2020nudge, wambsganss2022improving}. Concretely, our system analyzes the quality and type of feedback as soon as it is entered and provides visual indicators on a dashboard, enabling users to immediately recognize how their feedback might be improved. Second, the system provides the feedback recipient's (in this case, the AI mentee's) reactions—such as shifts in facial expression, inner thoughts, and an evolving knowledge level—according to the feedback. This design choice helps users understand how others receive their feedback, so they can reflect on how they can improve to better deliver their feedback to others~\cite{shaikh2024rehearsal, yeo2024help, kiskola2021applying}. Lastly, we designed the AI mentee to ask counter-questions, prompting users to offer feedback that they had not initially considered~\cite{cook2019guiding}. Specifically, we designed our system to analyze what was lacking in the user’s feedback and ask questions that could elicit that feedback from users.

Collectively, we refer to these system components as \textbf{Feedback Reflection Interface (FRI)} detailed in Section~\ref{sec:intervention_feature}. By designing the system around these elements, we aim to ensure that users experience a robust, realistic practice environment in which they can continually assess and refine their feedback skills. Note that we iteratively improved the \sysname{}'s interface and underlying pipeline based on pilot sessions with three design experts, each with over ten years of experience in design and design education.

\subsection{Evaluation of Design Feedback Qualities}  \label{sec:evaluation_criteria}

\begin{table*}
\centering
\scriptsize
\renewcommand{\arraystretch}{1.2}
\begin{tabular}{c|c|p{7.2cm}}
    \hline
    \multicolumn{2}{c|}{\textbf{Category}} & \textbf{Description} \\
    \hline
    \multirow{6}{*}{Question} 
        & \multirow{2}{*}{\centering Low-level Question} & The feedback leading to primary clarification of missing or incomplete information during communication.\\ \cline{2-3}
        & \multirow{2}{*}{\centering Deep Reasoning Question} & The feedback leading to causal explanations of the phenomenon under discussion.\\ \cline{2-3}
        & \multirow{2}{*}{\centering Generative Design Question} & The feedback leading to reframing and conceptual exploration of problem- and solution-spaces.\\
    \hline
    \multirow{6}{*}{Statement} 
        & \multirow{2}{*}{\centering Share Information} & The feedback that consists of additional information necessary to make progress on the task.\\ \cline{2-3}
        & \multirow{2}{*}{\centering Evaluation} & The feedback that assesses the quality of an individual answer or solution to the task.\\ \cline{2-3}
        & \multirow{2}{*}{\centering Recommendation} & The feedback that contains suggestions on how to improve the solution.\\
    \hline
\end{tabular}
\caption{Feedback typology used in \sysname{}.}
\label{tab:Feedback_Typology}
\end{table*}

\begin{table*}[t]
\centering
\scriptsize
\renewcommand{\arraystretch}{1.15}
\begin{NiceTabular}{%
  >{\centering\arraybackslash}p{1.5cm}|%
  >{\centering\arraybackslash}p{1.5cm}|%
  >{\centering\arraybackslash}p{1.9cm}|%
  p{7cm}}
\toprule
\multicolumn{2}{c}{\textbf{Feedback Type}} & \textbf{Criteria} & \textbf{Description}\\
\midrule
\multirow{17}{1.5cm}{\centering Single-turn}
  & \multirow{9}{1.5cm}{\centering Question}
      & \multirow{2}{1.9cm}{\centering Timeliness} & The feedback is timely, offering questions that match the recipient's current stage in the design process. \\ \cline{3-4}
  &  & \multirow{3}{1.9cm}{\centering Goal Relevance} & The feedback is aligned with the design goal and does not address points irrelevant to the recipient's hand-in.\\ \cline{3-4}
  &  & \multirow{3}{1.9cm}{\centering Level} & The feedback is appropriately challenging, requiring a degree of complex, critical, or creative thinking appropriate for the recipient.\\ \cline{3-4}
  &  & Sentiment & The feedback is positive (or negative) in tone.\\ \cmidrule(lr){2-4}
  & \multirow{7}{1.5cm}{\centering Statement}
      & \multirow{2}{1.9cm}{\centering Specificity} & The feedback is specific, pointing to exact design elements or artifacts.\\ \cline{3-4}
  &  & \multirow{2}{1.9cm}{\centering Justification} & The feedback is well-justified, backed by clear reasoning or evidence.\\ \cline{3-4}
  &  & \multirow{2}{1.9cm}{\centering Action} & The feedback is actionable and can be implemented immediately.\\ \cline{3-4}
  &  & Sentiment & The feedback is positive (or negative) in tone.\\
\midrule
\Block{2-2}{\makecell[c]{Multiple-turn}}
    && Ratio of Divergent and Convergent & The feedback session adapts to each state of the work, offering divergent feedback when exploration is needed and convergent feedback when focus is required.\\ \cline{3-4}
    && \multirow{4}{1.9cm}{\centering Ratio of Question and Statement} & The feedback session adapts to each state of the work, providing questions when open inquiry is appropriate and statements when definitive direction or consolidation is needed.\\
\bottomrule
\end{NiceTabular}
\caption{Evaluation criteria used in \sysname{}. Single-turn metrics assess the quality of an individual feedback turn, whereas multiple-turn metrics assess feedback at the session level.}
\label{tab:Feedback_Criteria}
\end{table*}

To design a system that aligns with our design rationales, we must clearly define what constitutes good feedback and how to evaluate it~\cite{ngoon2018interactive, yuan2016almost}. However, assessing feedback is inherently challenging because it is subjective and influenced by context and the recipient's perception~\cite{rucker2003assessing, yoshida2008teachers, cho2006commenting}. To address these challenges and evaluate feedback more objectively, prior research has shifted toward focusing on semantic and linguistic features rather than relying on subjective assessment from feedback recipients~\cite {cheng2020critique, cook2019guiding, hurst2019comparing}. Building on these approaches, we developed a feedback typology and corresponding criteria to implement in \sysname{}.

Since feedback varies in function and thus requires different evaluative criteria, it is important first to categorize the types of feedback to ensure accurate evaluation~\cite{hurst2019comparing}. Past research commonly breaks general feedback down into \stat{information}, \stat{evaluation}, and \stat{recommendation} components~\cite{shute2008focus, narciss1999motivational,tohidi2006getting}. However, unlike statement-based feedback, the conversational feedback we focus on evolves over multiple interactions, requiring the inclusion of \textit{questions} as part of the feedback process~\cite{hurst2019comparing, cardoso2020reflective, cordova2021comparison}. Design questions that stimulate creative thinking are often classified using Eris's taxonomy~\cite{eris2004effective}, which includes \question{low-level}, \question{deep reasoning}, and \question{generative design questions}. Our study combines these classifications and organizes feedback into six distinct types (Table~\ref{tab:Feedback_Typology}).

Next, we established specific evaluation criteria for each feedback type by referencing prior feedback assessments (Table~\ref{tab:Feedback_Criteria}). While many studies commonly evaluate statement-based feedback using criteria such as \textit{specificity}, \textit{justification}, and \textit{action}~\cite{cheng2020critique, krause2017critique, ngoon2018interactive, sadler1989formative, cook2019guiding, misiejuk2021using}, assessment of feedback that takes the form of questions does not have a widely accepted standard. To address this gap, we reviewed prior research and conducted an expert workshop, which led us to identify three criteria for question-based feedback: \textit{timeliness}~\cite{yen2024give, eris2004effective, thurlings2013understanding}, \textit{goal relevance}~\cite{misiejuk2021using, thurlings2013understanding}, and \textit{level}~\cite{eris2004effective, seaman2011bloom}. Given that design is an iterative process involving phases of exploration (i.e., divergence) and refinement (i.e., convergence), it is crucial to provide appropriate feedback that aligns with the specific needs of each phase~\cite{eris2004effective, yilmaz2016feedback, lekschas2021ask, marbouti2019written}. Thus, we introduced a measure that determines whether the current feedback is diverging or converging in nature and signals this to the user. 
Additionally, many studies have shown that feedback quality of feedback is significantly impacted by its sentiment~\cite{krause2017critique, yuan2016almost, marbouti2019written, wu2021better}, leading to the inclusion of \textit{sentiment} in our quality measurement.

These feedback typology and evaluation criteria are used for real-time feedback evaluation through LLMs, which is applied to implement the mentee's knowledge state and FRI.

\subsection{Design of Mentee Persona AI}
\begin{figure} []
\begin{center}
    \includegraphics[width=\textwidth]{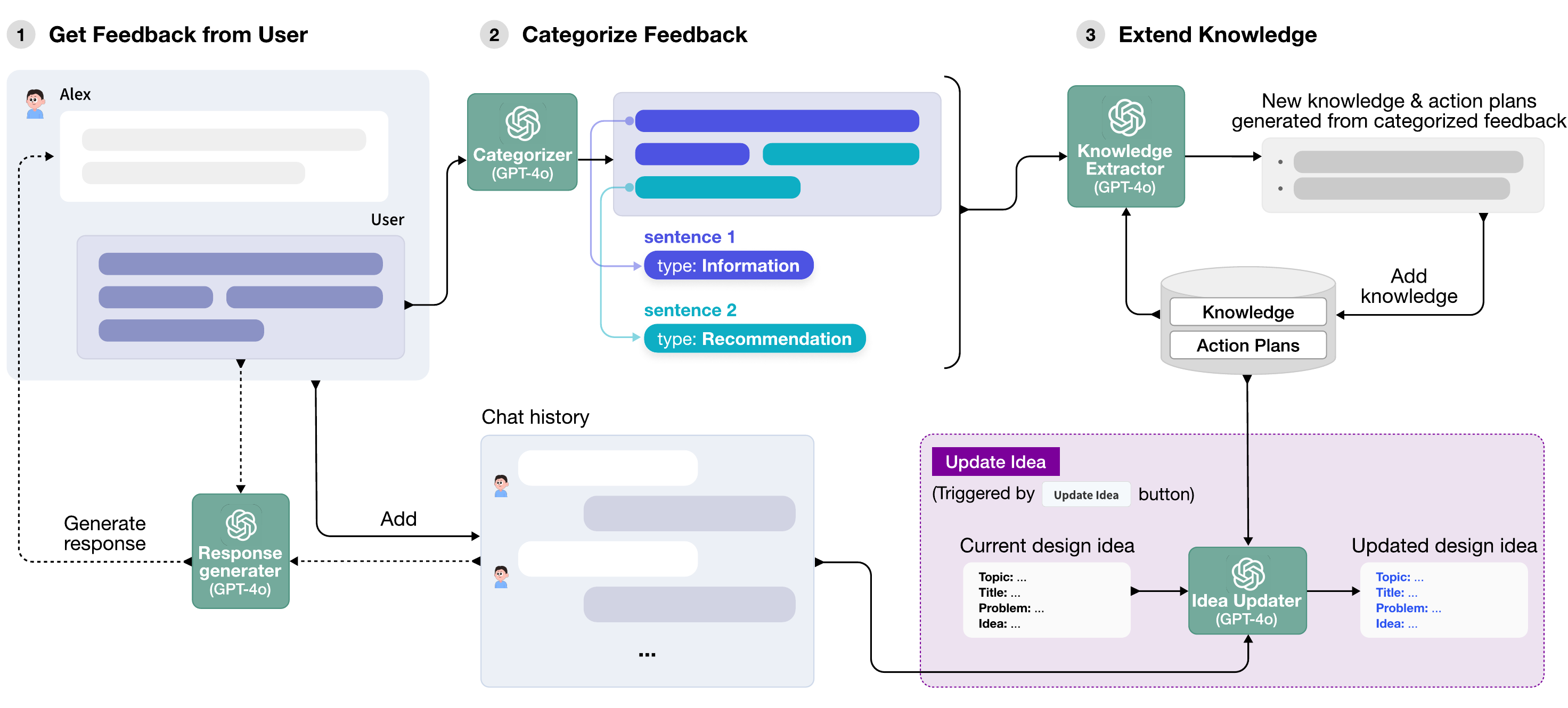}
    \caption[Structure of the Feed-O-Meter's baseline pipeline. First, feedback is provided by the user and processed by the categorizer, which is built using GPT-4o. The feedback is categorized into six predefined categories by the categorizer. Then, the knowledge extractor, created using GPT-4o, extracts knowledge and action plans from the categorized feedback and integrates them into the knowledge state. Subsequently, the response generator, also developed using GPT-4o, generates Alex’s responses to the user's feedback based on this information. When the user clicks the "Update Idea" button, the design idea is revised by the idea updater, which is likewise created using GPT-4o, according to the action plans and the chat history.]{Structure of the Feed-O-Meter's baseline pipeline. (1) Feedback is provided by the user and processed by the response generator through the following steps. (2) The categorizer categorizes the feedback into six predefined categories, such as information and recommendations. (3) \textit{knowledge} and \textit{action plan} are extracted by the knowledge extractor according to their categories and integrated into the knowledge state. When the user clicks the ``Update Idea'' button, a design idea is revised based on \textit{action plans} and the chat history.}
    \label{fig:baseline_pipeline}
\end{center}
\end{figure}

We developed an LLM-based agent named \mentee{} that functions as a mentee in role-playing scenarios. \mentee{} performs two key actions in role-playing: (1) responding to user feedback and (2) updating its ideas based on conversations with users.
This section provides a detailed overview of the technical aspects of these features, as illustrated in Figure \ref{fig:baseline_pipeline}, which outlines the entire pipeline.

\subsubsection{Simulating Immersive Feedback Interactions}
We established mentee personas not only through carefully crafted LLM prompts but also by embedding a knowledge state mechanism and conceptual design elements, thereby immersing users in role-playing and enhancing the reality of their responses~\cite{zhang2018personalizing}. Drawing on demographic information used in LLM-agent persona design~\cite{lee2025spectrum}, we added the social identity (name, nationality, and education level) and personal identity required for our role-playing context to the prompt. In sum, we modeled \mentee{} as a Korean first-year design major with limited design knowledge yet a strong desire for feedback on their project, mirroring a realistic scenario in which a novice seeks constructive input from experts. To bound the mentee's expertise, we initially assign an empty knowledge state, which is continuously updated throughout the feedback session. We also designed a cartoon-style portrait for the mentee featuring 25 distinct facial expressions to convey emotional states and deepen conversational immersion.

\mentee{} can interact with the user through a chat interface. \mentee{}'s responses are generated within 4-10 seconds, and during this time, a placeholder text such as \textit{``Umm...Uh...''} or \textit{``In my opinion...''} is displayed to simulate a natural conversation flow and maintain immersion.

\subsubsection{Shaping Ideas with User Feedback} \label{sec:updateidea}
We designed an LLM-based pipeline that updates \mentee{}'s design ideas based on the feedback provided by the user. It was important to ensure that these updates were confined to the knowledge \mentee{} could acquire through the conversation to maintain immersive experiences (DR1).

In detail, we first implemented an LLM-based categorizer (explained further in Section \ref{sec:feedbackeval}) to classify the user's feedback. When the feedback does not fall into categories such as ``no feedback'' or ``low-level question'' (which typically do not introduce new information or concepts), an LLM-powered knowledge extractor retrieves relevant \textit{knowledge} and \textit{action plans} from the feedback.
We defined \textit{knowledge} as high-level insights for the general design process, while the \textit{action plan} is specific guidance for the current design project. Therefore, we extracted and stored them separately and used the \textit{knowledge} to shape the ongoing conversation(e.g., prompting relevant counter-questions) and the \textit{action plan} to update the \mentee{}'s design idea.

\subsection{Design of Feedback Reflection Interface} \label{sec:intervention_feature}

\begin{figure} []
\begin{center}
    \includegraphics[width=\textwidth]{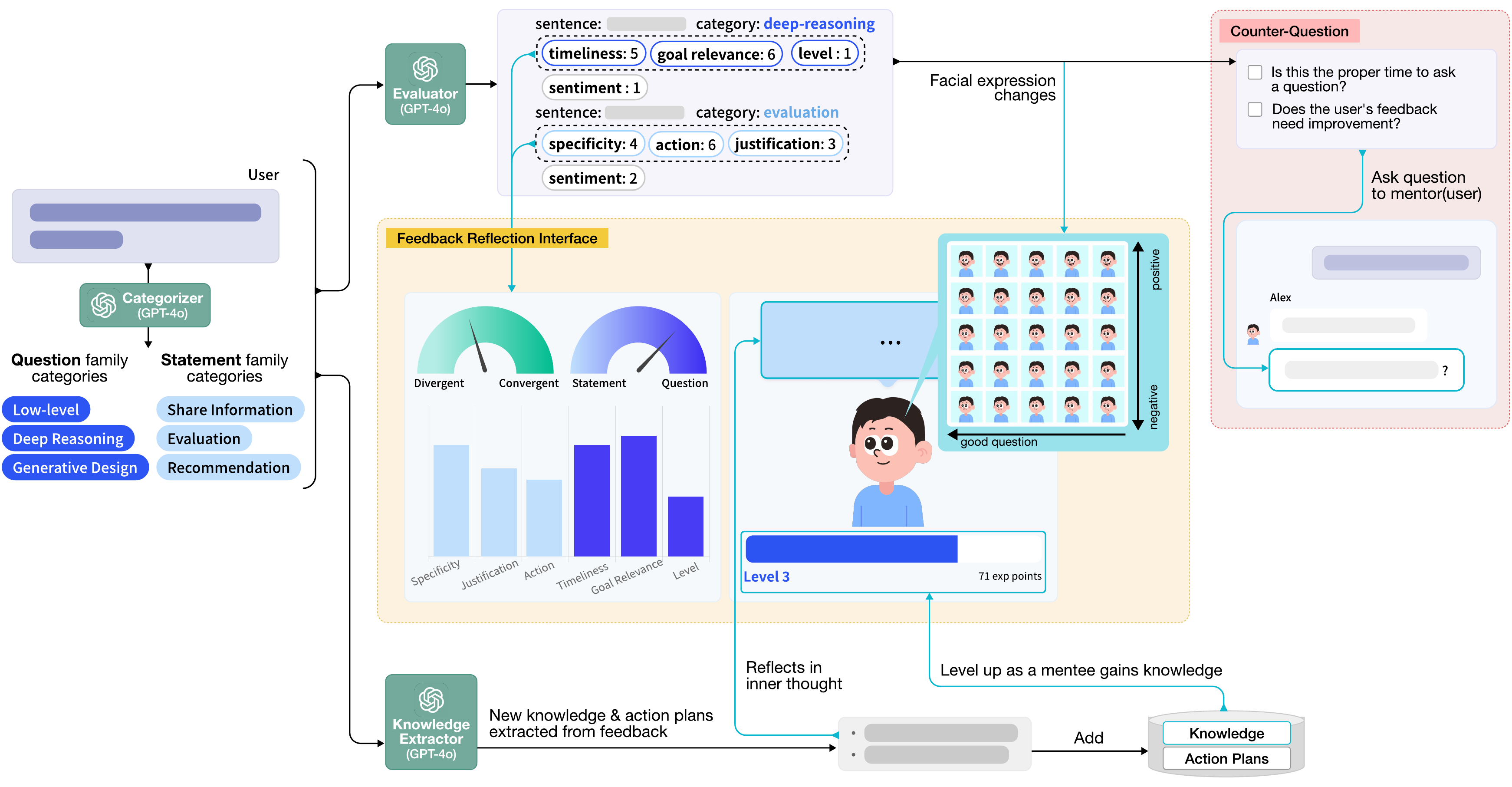}
    \caption[Pipeline of the feedback reflection interface. The pipeline begins by categorizing user feedback into one of six predefined categories—three from the question family (deep reasoning, evaluation, clarification) and three from the statement family (recommendation, share information, generative design). The categorization is handled by the categorizer, which is built using GPT-4o. Each feedback sentence is then evaluated based on criteria relevant to whether it belongs to the question family or the statement family by the evaluator, also developed using GPT-4o. The knowledge extractor, created using GPT-4o, extracts knowledge and action plans from the categorized feedback and accumulates them in the system. The newly generated knowledge is reflected in the mentee’s inner thoughts interface. These evaluation results are displayed in the feedback reflection interface, where the mentee’s facial expressions change dynamically in response to the quality and sentiment of the user's feedback. Counter-questions are generated by the system when two conditions are met: first, whether it is an appropriate time to ask a question, and second, whether there are missing elements in the user’s feedback. If both conditions are satisfied, a counter-question is generated.]{Pipeline of the Feedback Reflection Interface. The pipeline starts by categorizing user feedback into one of six categories—three from the question family and three from the statement family. Each feedback sentence is then evaluated according to criteria specific to its category. The evaluation results are displayed in the feedback reflection interface, influencing the mentee's facial expressions, which change dynamically based on the feedback. Counter-questions are generated when certain conditions are met.}
    \label{fig:intervention_pipeline}
\end{center}
\end{figure}

Beyond interacting with \mentee{}, \sysname{} incorporates interactive features that indirectly support users in improving their feedback during practice. The feedback reflection interface (Fig. \ref{fig:main_UI}-C) displays the results of feedback evaluations and \mentee{}'s reactions, allowing users to reflect on the quality and the impact of their feedback. Furthermore, \mentee{} poses counter questions to elicit more detailed feedback and thoughtful responses. This section explains the technical details of these features, and Figure ~\ref{fig:intervention_pipeline} illustrates the pipeline of these functionalities.

\subsubsection{Evaluation Dashboard} \label{sec:feedbackeval}

Building on the Feedback quality evaluation criteria from Section~\ref{sec:evaluation_criteria}, we developed a pipeline for evaluating feedback provided by users to \mentee{}. Since different feedback types require distinct evaluation criteria, we first implemented an LLM-based feedback categorizer. The categorizer classifies feedback into one of the six categories outlined in Table~\ref{tab:Feedback_Typology}, or returns ``No Feedback'' if the input lacks valid content.

Once categorized, the LLM-based evaluator applies specific criteria tailored to each feedback type, as specified in the Single-Turn column of Table~\ref{tab:Feedback_Criteria}, and determines where the feedback falls within these criteria. For questions, the evaluation focuses on timeliness (whether the feedback was posed at the appropriate time), goal relevance (whether it aligns with design goals), and level (high-level or low-level) of the question. For statements, the pipeline assesses specificity (how detailed the feedback is), justification (whether the feedback is supported by reasoning), and action (whether the feedback is actionable).  Additionally, the pipeline evaluates the sentiment of the feedback, determining whether it is a positive or negative tone. 

The results of this evaluation are updated in real-time on the dashboard, as shown in Figure~\ref{fig:main_UI}-C2. The dashboard displays the ratio of divergent and convergent feedback and the ratio of questions and statements, as well as each single-turn evaluation criteria. Sentiment, however, is not directly shown in the dashboard but is reflected through changes in the facial expression of the AI mentee.

\subsubsection{Mentee (\mentee{}) Profile: Visualizing Mentee's Reactions}

To enhance immersion (DR1) and support feedback improvement (DR2), we visualized \mentee{}'s reactions to the feedback provided by users (Fig~\ref{fig:main_UI}-C1). The interface offers three visualizations: Level Bar, Facial Expression, and Inner Thoughts. First, the Level Bar shows the amount of knowledge accumulated in \mentee{}'s knowledge state, encouraging users to provide more valuable feedback to enhance this level.

\mentee{}'s facial expressions are determined by two key factors: (1) the sentiment of the feedback and (2) the results from the single-turn evaluation. The sentiment factor is represented on a five-level scale---positive feedback makes \mentee{} smile, while negative feedback results in a sad expression. The second factor, the feedback evaluation, is also displayed on a five-level scale---highly evaluated feedback makes \mentee{}'s eyes sparkle, while poor feedback causes a skeptical expression on the eyes. By observing \mentee{}'s changing expressions, users are indirectly motivated to improve the quality of their feedback. 

We also introduce an interface displaying \mentee{}'s inner thoughts generated by LLMs based on the knowledge and action plans extracted from the idea-updating pipeline discussed in Section~\ref{sec:updateidea}. These inner thoughts are presented as concise, one-line sentences in a thought bubble above \mentee{}'s face.

\subsubsection{Asking Counter-Questions}
We designed a pipeline that allows \mentee{} to ask counter-questions when the user's feedback requires improvement or diversification. These counter-questions are triggered when users give repetitive feedback, such as consecutive questions or statements, prompting them to vary their responses. Also, counter-questions are generated when feedback evaluations show extremes, such as consistently low-level questions, or when feedback lacks specificity. In such cases, the system generates counter-questions to guide users toward better feedback. The system continuously tracks the types and quality of feedback throughout the conversation. We also set the threshold for triggering counter-questions for four consecutive occurrences of specific feedback conditions, as a lower threshold might disrupt the user's ability to provide feedback proactively.

\subsection{User Interface}
\subsubsection{Onboarding}
To help users fully embody the mentor’s role, we designed an onboarding interface where they create a mentor profile and set their feedback style and goals before entering the main interface (DR1). In this interface (Fig~\ref{fig:onboarding_UI}), users select one of five character profiles and answer three open-ended questions: `What type of mentor are you?', `What are the characteristics of your feedback?' and `What is the goal of the feedback session?'. This flow is designed to help users immerse themselves in the mentoring scenario by encouraging them to adopt a clear role and mindset before giving feedback.

\subsubsection{Main Interface}

The \sysname{} interface is designed as a chat-based web application (Fig~\ref{fig:main_UI}) to provide an intuitive and immersive experience for simulating feedback exchange scenarios. After completing the onboarding process, users are introduced to three main components: the Idea Proposal Interface (Fig \ref{fig:main_UI}-A), the Chat Interface (Fig~\ref{fig:main_UI}-B), and the Feedback Reflection Interface (Fig~\ref{fig:main_UI}-C).

The \textbf{Idea Proposal Interface} (left panel) displays the design project topic, design goals, and \mentee{}'s current design idea. \mentee{}’s idea includes a title, a description of the target problem, and an explanation of the proposed design solution. The ``Update Idea'' button (Fig~\ref{fig:main_UI}-A1) allows users to refresh and review updates to \mentee{}'s idea based on their ongoing conversation.

The \textbf{Chat Interface} (center) is where the interaction between the user and \mentee{} takes place. It begins with a message from \mentee{} saying, ``Hi, my name is \mentee{}. I appreciate any feedback on my idea.'' Users provide feedback through the chat while referring to \mentee{}’s current ideas displayed in the left panel, the chat history, and the feedback reflection interface on the right panel.

The \textbf{Feedback Reflection Interface} (right panel) helps users reflect on their feedback at a glance. It includes two non-interactive features: Mentee’s Profile (Fig~\ref{fig:main_UI}-C1) and the Feedback Analysis Dashboard (Fig~\ref{fig:main_UI}-C2). Mentee’s Profile displays \mentee{}'s facial expression reacting to feedback (Fig~\ref{fig:facial_expression}) and a ``thought bubble'' showing their inner thoughts. This affective feedback mechanism is designed based on related works, which showed that such expressions enhance user engagement and perceived social presence~\cite{brave2005computers}. The Feedback Analysis Dashboard provides real-time visualizations, including two O-Meters: one showing the ratio of divergent to convergent feedback and another displaying the ratio of question-based to statement-based feedback. A bar chart also presents the accumulated scores of each feedback criterion. 

\subsection{Implementation}
We developed the system interface using React\footnote{\href{https://react.dev/}{https://react.dev}} and connected it to a Flask\footnote{\href{https://flask.palletsprojects.com/}{https://flask.palletsprojects.com}}-based backend server that leverages the GPT API. We used OpenAI's chat completion API\footnote{\href{https://platform.openai.com/docs/guides/chat-completions}{https://platform.openai.com/docs/guides/chat-completions}} to analyze user feedback, generate mentee's responses, and update ideas. We employed the GPT-4o model, which is particularly suited for chat interactions due to its fast response generation speed. For parameter settings, we consistently used a temperature of 0, with all other values set to their defaults. All log data associated with each user is stored in a MySQL database. The source code of \sysname{} is publicly available at \url{https://github.com/Hyunseung-Lim/Feed-O-Meter}.

\subsection{Pipeline Evaluation}

\sysname{} incorporates various LLM-based modules, most of which utilize the LLM's text-generation capabilities. These capabilities are well-suited for tasks like generating responses to user feedback to enhance immersion. As these tasks represent well-established use cases for LLMs, we did not conduct a separate evaluation for these modules. However, the LLM-based feedback categorizer required a dedicated performance evaluation, as categorization accuracy can vary based on the prompts used. To confirm the reliability of the LLM-based categorizer, we used data from the user study. Out of 1,386 feedback sentences, we randomly selected 60, with 10 samples drawn from each of the six categories identified by the LLM pipeline. Two authors independently categorized these sentences, blinded to the LLM's results, and we measured the agreement between human and LLM classification without knowing the LLM's categorization results. Then, we measured the alignment between human and LLM classifications. The evaluation showed a Cohen's Kappa of 0.86 between the two authors, indicating strong agreement. When comparing the LLM's categorizations to the authors' labels, Cohen's Kappa values were 0.80 and 0.72, respectively. The results suggest substantial alignment between the LLM and human judgments, indicating that the LLM-based categorizer is both effective and reliable in accurately categorizing feedback. 

\section{User Study}
The user study investigates how \sysname{} influences users' design-feedback skills and their overall perceptions of the system. Rather than replacing existing peer feedback, \sysname{} introduces a novel experience in which students adopt the instructor’s perspective. Accordingly, our focus is not on direct comparisons with previous feedback practices but on how participants perceive this new approach to agent-based feedback training and how \sysname{} can be integrated into the current feedback practice. As outlined in the design rationales, \sysname{} aims to engage students in the practice of giving feedback and to help them reflect on their feedback to come up with higher-quality feedback. This leads us to the following research questions:

\begin{itemize}
    \item RQ1: How do students perceive the experience of providing feedback through \sysname{}?
    \item RQ2: How does the Feedback Reflection Interface (FRI), a set of novel features in our system, affect users' feedback skills?
\end{itemize}

To explore these questions, we conducted a within-group comparative study. This section provides a detailed description of our study design, which was approved by the university's Institutional Review Board (IRB).

\subsection{Participants}
We recruited 24 participants (12 females, 12 males) through online university communities in South Korea. The recruitment specifically targeted both undergraduate and graduate students majoring in design, with participants grouped by their design education experience in 2-year intervals to ensure a balanced distribution across experience levels. Demographic details are presented in \autoref{tab:demographic}. Participants' ages ranged from 19 to 32 ($M = 23.17$, $SD = 3.10$), with most majoring in Industrial Design and two participants majoring in Design \& Art. The study lasted 100 minutes, and participants were compensated 50,000 KRW (approximately 37 USD).


\begin{table*}[t]
    \centering 
    \renewcommand{\arraystretch}{1.3}
    {\scriptsize
    \begin{tabular}{cccccc}
        \toprule
        \textbf{\begin{tabular}[c]{@{}c@{}}Participant \\ ID\end{tabular}} &
        \textbf{\begin{tabular}[c]{@{}c@{}}Age\end{tabular}} &
        \textbf{\begin{tabular}[c]{@{}c@{}}Gender\end{tabular}} &
        \textbf{\begin{tabular}[c]{@{}c@{}}Experience in\\Learning Design\end{tabular}} &
        \textbf{\begin{tabular}[c]{@{}c@{}}Major\end{tabular}} &
        \textbf{\begin{tabular}[c]{@{}c@{}}Study \\ Condition\end{tabular}} \\ \toprule
        P1  & 20 & F & Under 2 years & Industrial Design & \multirow{12}{*}
        {\begin{tabular}[c]{@{}c@{}}\sysname{} \\ → \\ Baseline\end{tabular}} \\ \cline{1-5}
        P2  & 21 & F & 2 - 4 years & Industrial Design &  \\ \cline{1-5}
        P3  & 23 & F & 2 - 4 years & Design \& Art &  \\ \cline{1-5}
        P4  & 23 & F & 2 - 4 years & Industrial Design &  \\ \cline{1-5}
        P5  & 25 & F & 2 - 4 years & Design \& Art &  \\ \cline{1-5}
        P6  & 24 & F & Over 4 years & Industrial Design &  \\ \cline{1-5}
        P7  & 20 & M & Under 2 years & Industrial Design &  \\ \cline{1-5}
        P8  & 19 & M & Under 2 years & Industrial Design &  \\ \cline{1-5}
        P9  & 21 & M & 2 - 4 years & Industrial Design &  \\ \cline{1-5}
        P10 & 23 & M & 2 - 4 years & Industrial Design &  \\ \cline{1-5}
        P11 & 24 & M & Over 4 years & Industrial Design &  \\ \cline{1-5}
        P12 & 24 & M & Over 4 years & Industrial Design &  \\ \toprule
        P13 & 28 & F & Under 2 years & Industrial Design & \multirow{12}{*}
        {\begin{tabular}[c]{@{}c@{}}Baseline \\ → \\ \sysname{}\end{tabular}} \\ \cline{1-5}
        P14 & 22 & F & 2 - 4 years & Industrial Design &  \\ \cline{1-5}
        P15 & 23 & F & 2 - 4 years & Industrial Design &  \\ \cline{1-5}
        P16 & 23 & F & 2 - 4 years & Industrial Design &  \\ \cline{1-5}
        P17 & 24 & F & Over 4 years & Industrial Design &  \\ \cline{1-5}
        P18 & 32 & F & Over 4 years & Industrial Design &  \\ \cline{1-5}
        P19 & 19 & M & Under 2 years & Industrial Design &  \\ \cline{1-5}
        P20 & 18 & M & Under 2 years & Industrial Design &  \\ \cline{1-5}
        P21 & 25 & M & Under 2 years & Industrial Design &  \\ \cline{1-5}
        P22 & 25 & M & Over 4 years & Industrial Design &  \\ \cline{1-5}
        P23 & 23 & M & Over 4 years & Industrial Design &  \\ \cline{1-5}
        P24 & 27 & M & Over 4 years & Industrial Design &  \\ \bottomrule
    \end{tabular}
    }
    \vspace{0.2cm}
    \caption[The table presents the demographic information of 24 study participants. The columns include Participant ID, Age, Gender, Experience in Learning Design, Major, and Study Condition. Participants are divided into two groups. The first group, consisting of participants P1 through P12, completed the study in the Feed-O-Meter condition first, followed by the baseline condition. The second group, consisting of participants P13 through P24, completed the study in the baseline condition first, followed by the Feed-O-Meter condition. The ages of participants range from 18 to 32 years old, with varying experience in learning design, categorized as under 2 years, 2 to 4 years, and over 4 years. The majority of participants have a background in Industrial Design, with two participants (P3 and P5) majoring in Design and Art. The gender distribution is also noted, with a balanced representation of males and females.]{Demographic information of study participants. Study Condition refers to the order in which conditions were administered in the within-subject study.}
    \label{tab:demographic}
    \vspace{-0.5cm}
\end{table*}

\subsection{Study Design}
\ipstart{Study Conditions}
We conducted a within-subject study with two conditions: (1) a baseline condition and (2) the \sysname{} condition. In the baseline condition, participants used a version of \sysname{} without the Feedback Reflection Interface (FRI), meaning that the feedback evaluation was not displayed, and \mentee{} did not ask counter-questions. In the \sysname{} condition, all features of \sysname{} were activated. To maintain consistency in response time across both conditions, we used the same prompts to generate responses in the baseline condition, though the evaluation results were not displayed.

\ipstart{Tasks}
Participants engaged in a role-playing task in which they were asked to provide feedback on design ideas presented by an AI mentee. Their goal was to help the AI mentee refine the ideas and design objectives. The task was repeated three times: a 5-minute pre-session for exploring the interface and two 20-minute sessions under each condition (baseline and \sysname{} condition). While the task remained the same in each session, the AI mentee asked for feedback on different ideas each time.

Our system targeted the early ideation phase of the design process for problem-solving, a commonly adopted design task in the educational setting~\cite{Jonassen2000}. We selected three topics for the role-playing task--Carbon Emission Reduction, Pet Care, and Child Protection--based on example themes proposed by the renowned Red Dot Design Award\footnote{https://www.red-dot.org/design-concept/categories}. These socially, ethically, and environmentally relevant issues were chosen for their potential to broaden the focus of feedback beyond visual outcomes, encouraging critical discussions on how defined problems and generated ideas impact people and society. 
We also defined five design goals for the mentee's ideas--Innovation, Elaboration, Usability, Use Value, and Social Responsibility--based on criteria from the iF\footnote{https://ifdesign.com/en/} and Red Dot Awards.

We generated six seed design ideas for the AI mentee--two for each topic--that the mentee will present during the feedback sessions. To enhance realism, four undergraduate design students drafted ideas for each topic, including the title, target problem, and proposed solution. We selected six ideas from the initial 12 drafts by excluding overly abstract or narrowly specific ideas to allow substantive feedback sessions on each topic. The design ideas were presented in a different order for each participant, with researchers ensuring all ideas were used an equal number of times during the sessions.

\ipstart{Survey Design}
We asked participants to complete two types of surveys in our study. First, to assess whether \sysname{} impacts participants' feedback skills, we administered pre-surveys and post-surveys on feedback efficacy~\cite{tschannen2001teacher} to answer RQ2. Second, to compare the baseline and \sysname{} conditions, participants completed a debriefing survey after each experiment to answer RQ1. Based on our design rationales, we selected nine survey items: four regarding RQ1 and five regarding RQ2.

\subsection{Study Procedure}

\begin{figure*} []
\begin{center}
    \includegraphics[width=\textwidth]{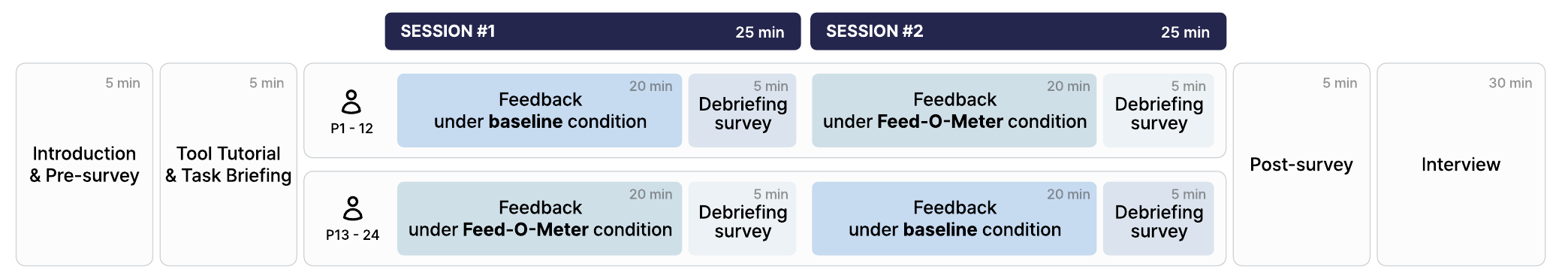}
    \caption[The figure outlines the study procedure, starting with a five-minute introduction and pre-survey. Following this, participants were given a five-minute tutorial to familiarize themselves with the tool and briefed on the task. The main study consisted of two sessions, each lasting 25 minutes. In one session, participants provided feedback under the baseline condition, and in the other session, they provided feedback under the Feed-O-Meter condition. Each session included a 20-minute feedback activity, followed by a five-minute post-session survey. After both sessions were completed, participants took a five-minute post-study survey on feedback efficacy. The study concluded with a 30-minute interview to gather qualitative insights into participants' experiences. The figure shows two groups of participants, with participants 1 to 12 completing the baseline condition first, while participants 13 to 24 started with the Feed-O-Meter condition.]{The outline of the user study with the time allocated for each step.}
    \label{fig:study_procedure}
    \vspace{-0.5cm}
\end{center}
\end{figure*}

The study was conducted in person, with an option for online participation via Zoom. The process is shown in \autoref{fig:study_procedure}. We first briefed the participants on the study's objectives and had them complete a pre-survey. After being introduced to \sysname{}'s interface and features, they had five minutes to explore the system and set up their mentor profiles (see \autoref{fig:onboarding_UI}). The study consisted of two 20-minute feedback sessions: one under the \sysname{} condition and one under the baseline condition. During these sessions, one was with \sysname{} and one in the baseline condition. During these sessions, participants provided feedback on design ideas while interacting with an AI mentee, clicking the ``Update Idea'' button at least once. After each session, participants completed a debriefing survey, and after both sessions, a post-survey on feedback efficacy. The study concluded with a 30-minute interview to gather insights into their experience and perceptions. The interview protocol covered the overall experience of using the system, the characteristics of participants' feedback with \sysname{} in comparison to real-world feedback scenarios, and the difference in the experience of providing feedback between the \sysname{} and baseline conditions.

\subsection{Measures and Analysis}
We collected log data of all interactions between users and \sysname{} and audio-recorded post-interview sessions. Both quantitative and qualitative methods were used to analyze the log data, survey responses, and interview transcripts. First, we performed descriptive statistical analysis on participants' feedback and \mentee{}'s responses. To gain deeper insights, we employed open coding along with thematic analysis~\cite{virginia2006thematic}. First, we divided one user’s feedback log into individual sentences because it contained multiple sentences featuring various feedback types. Two researchers then independently categorized these sentences into six pre-defined categories (see ~\autoref{tab:Feedback_Typology}), which were further refined into 15 subcategories through discussion (see ~\autoref{tab: Feedback_categorizing_result}).

Additionally, we conducted an expert evaluation of the log data to compare the quality of feedback in each condition. We recruited 12 design experts (eight males, four females; mean age = 29.33, $SD = 3.11$), each holding at least a master’s degree in industrial design, an average of 9.5 ($SD = 2.24$) years of experience, and having taught design at the college level. We randomly assigned eight feedback sessions (four \sysname{} conditions and four baseline conditions) to each expert, resulting in a total of 96 evaluated feedback sessions. For each session, experts assessed every sentence-level user log as well as the entire session according to specified criteria (see \autoref{tab:Feedback_Criteria}). Experts rated \textit{timeliness}, \textit{goal relevance}, and \textit{level} on a 7-point Likert scale for question-based feedback at the sentence level, and \textit{specificity}, \textit{justification}, and \textit{action} on a 7-point Likert scale for statement-based feedback. After reviewing all user logs, the experts provided overall ratings on a 7-point Likert scale based on three criteria--two from the typology in \autoref{tab:Feedback_Criteria} (\textit{ratio of divergent and convergent feedback}, \textit{ratio of question and statement feedback}) and \textit{overall helpfulness}. Finally, they were invited to leave open-ended comments.

We analyzed four sets of survey data. First, to compare participants’ feedback experience between the two sessions (\sysname{} and baseline), we conducted paired t-tests on each post-session survey item. We also reported effect sizes as Hedges’ g~\cite{hedges1981distribution} to contextualize statistical significance and mitigate concerns that our inferences are driven by sample size alone. Because these items were conceptually independent rather than forming a single family, we did not apply a Bonferroni correction. Second, to examine changes in feedback efficacy from pre- to post-experiment, we ran paired t-tests on the pre- and post-surveys and likewise reported Hedges’ g. In this case, to control for Type I errors across multiple comparisons, we applied a Bonferroni correction.

Finally, we conducted a thematic analysis~\cite{virginia2006thematic} of interview transcripts to complement the survey and log data findings. The first author performed open coding, and the research team identified overarching themes through discussion, adding depth and validity to our qualitative analysis.

\section{Findings}
In this section, we present the key findings from our study. First, we provide a descriptive summary outlining the details of the overall use of \sysname{}. Second, we compare the feedback quality and participants' usage experiences between the baseline and \sysname{} conditions to determine how our Feedback Reflection Interface (FRI) influenced their quality of feedback. Finally, we examine the participants' overall perspective on \sysname{} as a design feedback practice system.

\subsection{Descriptive Summary of \sysname{} Usage}
\subsubsection{Feedback Provided by Participants}

\begin{table*}[t]
    \scriptsize
    \setlength{\tabcolsep}{5pt}
    \begin{tabular}{lcccccc}
    \toprule
      & \multicolumn{2}{c}{\sysname{}} & \multicolumn{2}{c}{baseline} & \multicolumn{2}{c}{statics} \\ \cline{2-7}
      & mean & std & mean & std & P & sig. \\
    \toprule
    \# of (user's) feedback & 12.63 & 4.15 & 14.29 & 6.60 & 0.3005 & - \\
    \# of syllables in (user's) feedback & 207.85 & 135.99 & 183.11 & 138.94 & 0.0229 & $\ast$ \\
    \# of (user's) feedback (by sentence) & 25.25 & 6.15 & 26.41 & 9.15 & 0.6066 & - \\
    \# of syllables in (user's) feedback (by sentence) & 103.40 & 62.94 & 98.80 & 61.03 & 0.1924 & - \\
    \hline
    \# of (mentee's) responses &12.63 & 4.15 & 14.29 & 6.60 & 0.3005 & - \\
    \# of syllables in (mentee's) responses & 218.70 & 76.08 & 210.67 & 75.35 & 0.1788 & -\\
    \# of (mentee's) counter questions & 3.79 & 1.10 & - & - & - & - \\
    \# of syllables in (mentee's) counter questions & 132.79 & 37.94 & - & - & - & - \\
    \hline
    \# of clicks on Update Idea button & 2.63 & 1.06 & 2.91 & 1.56 & 0.4516 & - \\
    \# of syllables in (mentee's) idea & 1358.75 & 213.07 & 1356.89 & 187.70 & 0.9574 & - \\
    \bottomrule
    \end{tabular}
    \vspace{0.2cm}
    \caption[The table compares interaction metrics between baseline and Feed-O-Meter conditions, showing means, standard deviations, p-values, and significance levels. Key metrics include the number of user feedback, mentee responses, syllables in feedback and responses, counter-questions (only in Feed-O-Meter), and clicks on the "Update Idea" button. The Feed-O-Meter condition shows slightly more syllables in user feedback and mentee counter-questions, with a statistically significant difference in the number of syllables in user feedback (p = 0.0229). Most other metrics, including the number of feedback, responses, and idea syllables, show no significant differences between conditions.]{Statistical summary of interaction logs (include the number of user feedback, mentee responses, syllables in feedback and responses, counter-questions (only in Feed-O-Meter), and clicks on the Update Idea button) between \sysname{} and baseline conditions. (- : $p > .05$, $\ast$ : $p < .050$, $\ast\ast$ : $p < .010$, $\ast\ast \ast$ : $p < .001$)}
    \label{tab:statistical_summary}
    \vspace{-0.5cm}
\end{table*}

In the user study, 24 participants each completed two feedback sessions, exchanging a total of 1,431 chats with the AI mentee (of which 646 were from the participants). In the \sysname{} condition, they provided an average of 12.63 ($SD = 4.15, min = 7 [P4], max = 22 [P1]$) chats, with each chat averaging 207.85 ($SD = 135.99, min = 78.38 [P3], 471.86 [P4]$) syllables per chat. In the baseline condition, participants provided an average of 14.29 ($SD = 6.60, min = 5 [P2], max = 34 [P1]$) chats over 20 minutes, with each chat averaging 183.11 ($SD = 138.94, min = 93.68 [P3], max = 453.60 [P2]$) syllables per chat. Participants' feedback was statistically significantly longer in the \sysname{} condition ($t = -2.28, p = 0.0229$).

\begin{table}[]
\centering
\scriptsize
\renewcommand{\arraystretch}{1.3}
\resizebox{\columnwidth}{!}{%
\begin{tabular}{p{2cm}p{2.5cm}|p{4cm}|p{6cm}}
\toprule
\textbf{Category} & \textbf{Sub-Category} & \textbf{Description} & \textbf{Example} \\
\toprule
\multirow{7}{=}[-0.5ex]{Low-Level \\ (F: 81, B: 114)} & Verification \newline (F: 9, B: 18) & Feedback to make sure the user understands the mentee's idea. & \textbf{P22}: So this idea is a filtering service for kids? \newline \textbf{\mentee{}}: Yes, that's right.\\
\cline{2-4}
 & Completion \newline (F: 43, B: 76) & Feedback to clarify something that is not clearly explained. & \textbf{P13}: How exactly does virtual adoption work? \newline \textbf{\mentee{}}: It uses VR to simulate pet ownership experience.\\
\cline{2-4}
 & Understanding Mentee \newline (F: 23, B: 15) & Feedback to get to know mentee's background, understanding, interests, and more. & \textbf{P21}: \mentee{}, do you have any pets? \newline \textbf{\mentee{}}: Yes, my family has a dog, and that's what inspired me to come up with this idea.\\
\hline
\multirow{7}{=}[-0.5ex]{Deep Reasoning \\ (F: 46, B: 72)} & Logical / Causal \newline Reasoning \newline (F: 37, B: 60) & Feedback that prompts the mentee to reason about the feasibility, realization, effectiveness, etc. of the idea. & \textbf{P24}: Is it scientifically possible? \newline \textbf{\mentee{}}: To be honest, I haven't thought deeply about that.\\
\cline{2-4}
 & Instrumental / Procedural Reasoning \newline (F: 8, B: 11) & Feedback asking about the procedure and reasons behind the mentee's decision. & \textbf{P15}: Why did you limit the target to children under 7 years old? \newline \textbf{\mentee{}}: Oh, I didn't limit it to children under 7.\\
\hline
\multirow{8}{=}[-0.5ex]{Generative Design \\ (F: 35, B: 57)} & Brainstorming /\newline Ideation \newline (F: 19, B: 26) & Feedback that provides or elicits ideas without a deliberate end goal. & \textbf{P15}: How about letting them know in the dog's voice saying "I want to go for a walk"?\\
\cline{2-4}
 & Negotiation \newline (F: 7, B: 16) & Feedback to suggest/negotiate the new idea instead of the current one. & \textbf{P19}: Is there any way we could detect child abuse earlier, before it gets too serious? \\
\cline{2-4}
 & Scenario Creation \newline (F: 9, B: 12) & Feedback that presents specific scenarios that could happen. & \textbf{P20}: In abusive households, parents might prevent children from making emergency calls. How can we address this issue?\\
\hline
\multirow{6}{=}[-0.5ex]{Share Information \\ (F: 75, B: 46)} & Sharing Examples / Personal Experience \newline (F: 34, B: 26) & Feedback that provides an example or personal experience & \textbf{P10}: Have you heard of `Elsagate'? [...], it seems difficult to filter out malicious content similar to those interests.\\
\cline{2-4}
 & Providing Design \newline Knowledge \newline (F: 24, B: 15) & Feedback that provides design knowledge or principles. & \textbf{P6}: Another important factor to consider is what stakeholders can help when child abuse issues occur. [...] It's important to consider these various stakeholders.\\
\hline
\multirow{6}{=}[-0.5ex]{Evaluation \\ (F: 137, \newline B: 162)} & Positive\newline Assessment \newline (F: 63, B: 67) & Feedback that explicit positive assessment of the quality of the design. & \textbf{\mentee{}}: Users could express satisfaction with emoticons after watching content. \newline \textbf{P16}: Oh, using emoticons for feedback is a great idea!\\
\cline{2-4}
 & Negative\newline Assessment \newline (F: 58, B: 85) & Feedback that explicit negative assessment of the quality of the design. & \textbf{P10}: I got the impression that the target problem and the ideas aren't really well aligned..\\
\hline
\multirow{8}{=}[-0.5ex]{Recommendation \\ (F: 193, \newline B: 144)} & Direct\newline Recommendation \newline (F: 103, B: 86) & Feedback that gives specific advice on what or how to do. & \textbf{P23}: Let's design a platform that's not just for adopters, but one that various stakeholders from each facility can use together.\\
\cline{2-4}
 & Hinting \newline (F: 73, B: 50) & Feedback that indirectly suggests a way to proceed without making a direct suggestion. & \textbf{P9}: You should look that up. As a hint, think about what's currently used in automatic doors.\\
\cline{2-4}
 & Project\newline Management \newline (F: 15, B: 8) & Feedback on project management, including scheduling, deliverables, and stakeholder management. & \textbf{P18}: It would be good to organize the types and situations of child abuse by third parties indoors more specifically.\\
 \hline
\multirow{3}{=}[-0.5ex]{No Feedback \\ (F: 39, B: 39)} & \multirow{2}{=}{} & \multirow{2}{=}{Social expression, empathy, and compliments} & \textbf{P4}: Hello, I've carefully read your ideas. \\
\cline{4-4}
 &  &  & \textbf{P5}: I didn't give you much advice, but you're really good at developing ideas! Haha.\\
\bottomrule
\end{tabular}
}
\vspace{0.2cm}
\caption[The table categorizes user feedback into several subtypes under different categories, comparing the counts between the Baseline (B) and \sysname{} (F) conditions. Each sub-category includes a description and an example. Under the Low-Level category (B: 114, F: 81), the sub-categories are Verification (B: 18, F: 9), Completion (B: 76, F: 43), and Understanding Mentee (B: 15, F: 23). For Deep Reasoning (B: 72, F: 46), the sub-categories are Logical/Causal Reasoning (B: 60, F: 37) and Instrumental/Procedural Reasoning (B: 11, F: 8). Under Generative Design (B: 57, F: 35), the sub-categories are Brainstorming/Ideation (B: 26, F: 19), Negotiation (B: 16, F: 7), and Scenario Creation (B: 12, F: 9). For Share Information (B: 46, F: 75), the sub-categories are Sharing Examples/Personal Experience (B: 26, F: 34) and Providing Design Knowledge (B: 15, F: 24). The Evaluation category (B: 162, F: 137) has two sub-categories: Positive Assessment (B: 67, F: 63) and Negative Assessment (B: 85, F: 58). Under Recommendation (B: 144, F: 193), the sub-categories are Direct Recommendation (B: 86, F: 103), Hinting (B: 50, F: 73), and Project Management (B: 8, F: 15). Finally, under No Feedback (B: 39, F: 39), there are no sub-categories, but it includes social expressions, empathy, and compliments.]{Categorization of user feedback at the sentence level. Six predefined categories have 15 subcategories, each with associated descriptions and examples. The number of feedback instances belonging to the \sysname{} condition is indicated by F, and those in the Baseline condition are indicated by B.}
\label{tab: Feedback_categorizing_result}
\vspace{-0.5cm}
\end{table}

Participants provided various types of feedback to improve \mentee{}'s ideas (see ~\autoref{tab: Feedback_categorizing_result}). At the sentence level, the most frequent type of question-based feedback was \question{Low-Level Questions} ($N=195$), followed by \question{Deep Reasoning} \question{Questions} ($N=118$) and \question{Generative Design Questions} ($N=92$). Interestingly, the total number of question-based feedback was higher in the baseline condition ($N=243$) compared to the \sysname{} condition ($N=162$) across all sub-categories except for the Understanding Mentee, where the \sysname{} condition ($N=23$) had more questions than the baseline ($N=15$). Among statement-based feedback, \stat{Recommendation} ($N=337$) were the most common, followed by \stat{Evaluation} ($N=299$) and \stat{Sharing Information} ($N=121$). 
While statement-based feedback appeared more frequently in the \sysname{} condition ($N=405$) than in the baseline condition ($N=352$), \stat{Evaluation} was more prevalent in the baseline condition ($N=162$) compared to the \sysname{} condition ($N=137$). Notably, in the \sysname{} condition, Positive Assessment ($N=63$) was more frequent than Negative Assessment ($N=58$), whereas, in the baseline condition, Negative Assessment ($N=85$) appeared more often than Positive Assessment ($N=67$).

\subsubsection{AI Mentee's Responses to Feedback}
A total of 737 responses from \mentee{} were observed, excluding the starting messages. Of these, 646 were direct responses to user feedback, and 91 were counter-questions generated only in \sysname{} condition. The average response length was 218.70 ($SD = 76.08$) syllables in the \sysname{} condition and 210.67 ($SD = 75.35$) syllables in the baseline condition. 

\subsubsection{Ideas generated by AI Mentee}
Participants requested a total of 133 idea revisions by clicking the ``Update Idea'' button during the feedback sessions, averaging 2.63 times ($SD = 1.06$) in \sysname{} condition and 2.92 times ($SD = 1.56$) in the baseline condition. A total of 48 ideas were generated across three topics, with 16 ideas per topic. The average idea length was 1310.50 ($SD = 341.00$) syllables in the initial idea, 1358.75 ($SD = 213.07$) syllables in the \sysname{} condition, and 1356.89 ($SD = 187.70$) syllables in the baseline condition. A t-test showed no statistically significant difference in idea length between the two conditions ($t = -0.05, p = 0.9574$).

\subsection{Impact of the Feedback Reflection Interface on Feedback}
To answer RQ2, we analyzed the participants' feedback quality during the study session and their perception difference between the two conditions. In this session, we will first report on the difference in feedback quality in each condition, followed by the difference in participants' feedback experiences in each condition.

\begin{figure*} []
\begin{center}
    \includegraphics[width=\textwidth]{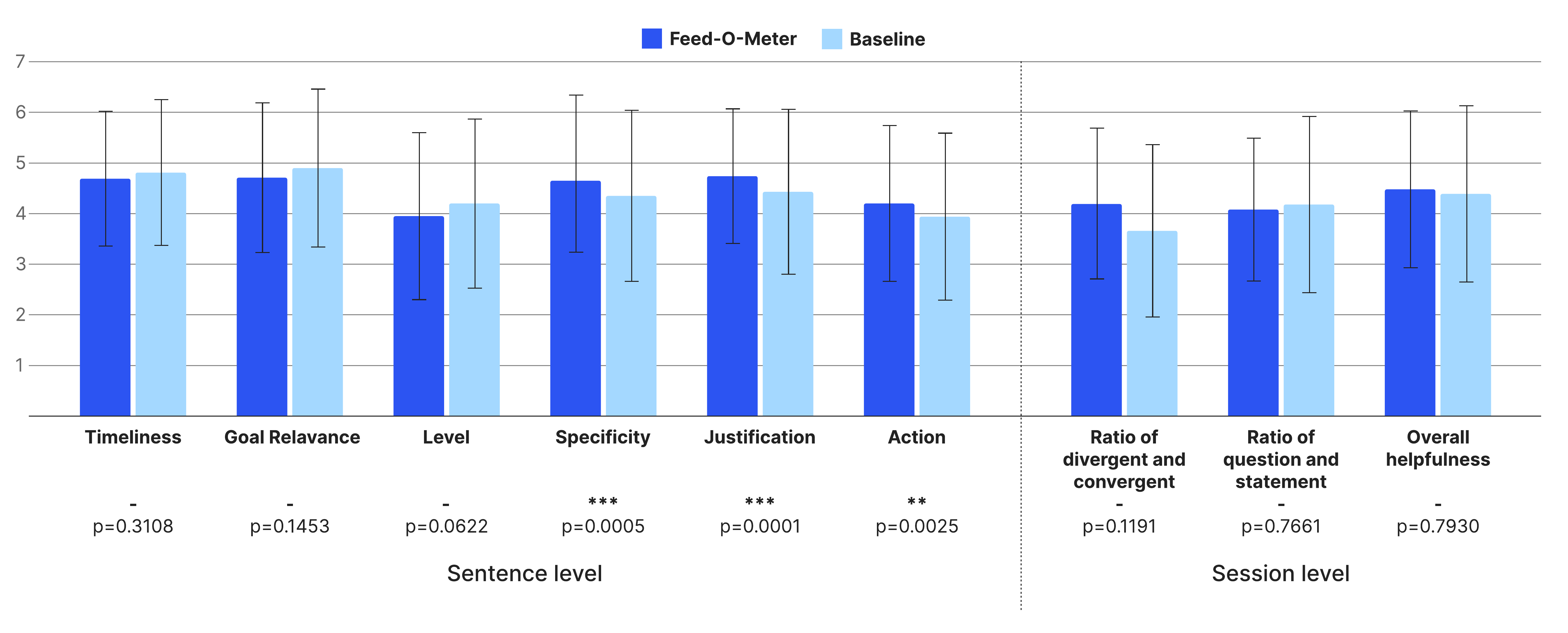}
    \caption[]{Comparison of expert evaluation of participants' feedback under the \sysname{} condition and the baseline condition. For sentence-level, question-based feedback was evaluated by \textit{timeliness}, \textit{goal relevance}, and \textit{level}, while statement-based feedback was evaluated by \textit{specificity}, \textit{justification}, and \textit{action}. The overall feedback session was evaluated by \textit{ratio of divergent and convergent}, \textit{ratio of question and statement}, and \textit{overall helpfulness}. (- : $p > .05$, $\ast$ : $p < .050$, $\ast\ast$ : $p < .010$, $\ast\ast \ast$ : $p < .001$)}
    \label{fig:evaluation_result}
    \vspace{-0.5cm}
\end{center}
\end{figure*}

\subsubsection{Feedback quality between \sysname{} and baseline conditions}
\autoref{fig:evaluation_result} shows the expert evaluation results comparing the user experience of \sysname{} with the FRI and the baseline system.
For the question-based feedback criteria, \textit{timeliness} (\sysname{}: $M = 4.68, SD = 1.34$ / baseline: $M = 4.80, SD = 1.45$ / $p = 0.3108$), \textit{goal relevance} (\sysname{}: $M = 4.70, SD = 1.48$ / baseline: $M = 4.89, SD = 1.57$ / $p = 0.1453$), and \textit{level} (\sysname{}: $M = 3.94, SD = 1.66$ / baseline: $M = 4.19, SD = 1.68$ / $p = 0.0622$) all showed no statistically significant differences between the two conditions. Meanwhile, for the statement-based feedback criteria of \textit{specificity} (\sysname{}: $M = 4.64, SD = 1.56$ / baseline: $M = 4.34, SD = 1.70$ / $p = 0.0005$), \textit{justification} (\sysname{}: $M = 4.73, SD = 1.43$ / baseline: $M = 4.42, SD = 1.64$ / $p = 0.0001$), and \textit{action} (\sysname{}: $M = 4.19, SD = 1.55$ / baseline: $M = 3.93, SD = 1.66$ / $p = 0.0025$), the \sysname{} condition showed significantly higher scores across all measures. For the overall evaluation of the feedback session, the \textit{ratio of divergent and convergent} (\sysname{}: $M = 4.18, SD = 1.50$ / baseline: $M = 3.65, SD = 1.71$ / $p = 0.1191$) was higher in the \sysname{} condition but did not reach statistical significance, and neither the \textit{ratio of question and statement} (\sysname{}: $M = 4.07, SD = 1.42$ / baseline: $M = 4.17, SD = 1.75$ / $p = 0.7661$) nor \textit{overall helpfulness} (\sysname{}: $M = 4.47, SD = 1.56$ / baseline: $M = 4.38, SD = 1.75$ / $p = 0.7930$) showed a statistically significant difference.

\subsubsection{Participants' experiences between \sysname{} and baseline conditions}

\begin{table*}[t]
    \scriptsize
    \setlength{\tabcolsep}{5pt}
    \begin{tabular}{p{5.8cm}cccccc}
    \toprule
    & \multicolumn{2}{c}{\sysname{}} & \multicolumn{2}{c}{baseline} & \multicolumn{2}{c}{statics} \\ \cline{2-7}
    & mean & std & mean & std & $P$ (sig.) & ES \\
    \toprule
    It was enjoyable to provide feedback through the system. & 6.54 & 0.59 & 5.88 & 0.95 & $0.0053\,(\ast\ast)$ & 0.82 \\
    I would like to practice giving feedback through the system in the future. & 6.08 & 0.97 & 5.71 & 0.95 & $0.1846\,(-)$ & 0.38 \\
    Giving feedback on design ideas through the system felt similar to giving feedback on real design projects. & 5.62 & 1.38 & 5.50 & 1.56 & $0.7699\,(-)$ & 0.08 \\
    The mentee (\mentee{}) resembled real design students. & 5.12 & 1.30 & 4.58 & 1.56 & $0.1969\,(-)$ & 0.37 \\
    \hline
    The system helped me recognize whether I was providing good feedback. & 6.00 & 0.88 & 3.75 & 1.15 & $0.0000\,(\ast\ast\ast)$ & 2.16 \\
    Conversations with the mentee (\mentee{}) inspired me to provide feedback I hadn't considered before. & 5.83 & 1.17 & 4.79 & 1.67 & $0.0158\,(\ast)$ & 0.53 \\
    It was easy to provide feedback through the system. & 5.42 & 1.28 & 4.54 & 1.69 & $0.0495\,(\ast)$ & 0.57 \\
    Using the system has enhanced my feedback skills. & 5.62 & 1.06 & 5.04 & 1.20 & $0.0799\,(-)$ & 0.50 \\
    My feedback improved the mentee (\mentee{})'s idea. & 6.25 & 0.53 & 5.58 & 1.21 & $0.0174\,(\ast)$ & 0.71 \\
    \bottomrule
    \end{tabular}
    \vspace{0.2cm}
    \caption[The table presents results from a debriefing survey with nine questions, comparing user experiences between the \sysname{} and baseline conditions. For each question, the p-value (P) and significance level (sig.) are shown. Users found providing feedback significantly more enjoyable in the \sysname{} condition (p = 0.0053). While participants expressed interest in using the system in the future, this was not statistically significant. The system significantly helped participants recognize whether they were providing good feedback (p = 0.0000). Conversations with the mentee in the \sysname{} condition inspired new types of feedback (p = 0.0158), and it was easier to provide feedback compared to the baseline (p = 0.0495). Additionally, users believed their feedback improved the mentee's idea more in the \sysname{} condition (p = 0.0174).]{Nine themed questions were given in a debriefing survey. The four questions above are related to RQ1, and the five below are related to RQ2 (- : $p > .05$, $\ast$ : $p < .050$, $\ast\ast$ : $p < .010$, $\ast\ast \ast$ : $p < .001$). Effect sizes(ES) are reported as \textit{Hedges’ $g$} (pooled-mean SD standardization with small-sample correction). 
    Magnitudes: small $\approx 0.20$, medium $\approx 0.50$, large $\approx 0.80$.}
    \label{tab:survey_of_intervention}
    \vspace{-0.5cm}
\end{table*}

\autoref{tab:survey_of_intervention} shows the post-survey results comparing the user experience of \sysname{} with the FRI and the baseline system. Participants found providing feedback with \sysname{} more enjoyable (baseline: $M = 5.88, SD = 0.95$ / \sysname{}: $M = 6.54, SD = 0.59$ / $p = 0.0053^{**}$). However, there was no significant difference in whether the AI mentee felt more like a design student.
Participants indicated that the FRI helped participants recognize whether they were giving good feedback (\sysname{}: $M=6.00, SD=0.88$ / baseline: $M=3.75, SD=1.15$ / $p=0.0000^{***}$), and counter-questions inspired them to provide more feedback (\sysname{}: $M=5.83, SD=1.17$ / baseline: $M=4.79, SD=1.67$ / $p=0.0158^{*}$). Participants also felt that \mentee{}’s ideas improved more with feedback through the \sysname{} (\sysname{}: $M=6.25, SD=0.53$ / baseline: $M=5.58, SD=1.21$ / $p=0.0174^{*}$) and found feedback easier to provide (\sysname{}: $mean = 5.42, SD = 1.28$ / baseline: $mean = 4.54, SD = 1.69$ / $p = 0.0495^{*}$). However, there was no significant difference in perceived improvement in their feedback skills (\sysname{}: $M=5.62, SD=1.06$ / baseline: $M=5.04, SD=1.20$ / $p=0.0799$). 

To gain deeper insights into participants' experiences with the two conditions, we identified key themes from their subjective perspectives and experiences through qualitative analysis.

\ipstart{Enhancing Feedback Clarity for the Mentee}
Most participants (22/24) found that the visualized information on the FRI helped them reflect on their feedback patterns and how \mentee{} perceived them. For example, P18 mentioned, \textit{``I enjoyed seeing the \mentee{}'s inner thoughts to my feedback. It's a perspective we don't usually have, so it was insightful and engaging.''} For some participants (6/24) with less feedback experience, this visualization served as a guide for providing constructive feedback. P8 said, \textit{``The interface showed divergent and convergent feedback, right? I realized feedback has these criteria and that most of mine was convergent.''} 

The visual nature of FRI allowed participants to check if their feedback was delivered as intended and adjust it for clarity when necessary. P19 stated, \textit{``It was easier to give feedback when I could see how it was understood in the top section (Mentee's Profile). I would rephrase it if \mentee{} did not get my point.''} Participants also adjusted the flow or style of their feedback. P18 shared, \textit{``I initially tried to give more divergent feedback but realized most of what I said was actually convergent. So, I started asking things like, `What do you think?' instead of just giving definitive answers.''} Although participants recognized their feedback patterns through the FRI, not all adjusted their feedback. Some (5/24) found it challenging to act on the information. For example, P3 wanted to improve a specific metric but struggled with how to do it. Likewise, P4 mentioned, \textit{``The criteria were kind of abstract, so it was hard to tell if I was doing well.''}

\ipstart{Diversifying Feedback through Counter-Questions}

Some participants (10/24) appreciated how counter-questions helped break repetitive feedback patterns and pushed the discussion in deeper, more diverse directions. P1 observed that \mentee{}'s questions introduced new topics, moving the conversation into new phases. P23 shared, \textit{``When I thought there was nothing left to give feedback on, \mentee{} suddenly asked what I thought about the financial aspects. I hadn't expected that.''} Some participants (8/24) enjoyed engaging with diverse perspectives. P10 remarked, \textit{``Since it's a design project, I can't just make everything to fit my preferences. In the \sysname{} condition, \mentee{}'s differing views encouraged me to broaden my thinking and find a middle ground.''}

\ipstart{Motivating Feedback Engagement}
Half of the participants (12/24) reported feeling a sense of accomplishment when their experience points increased or when \mentee{} displayed a happy facial expression. P15 noted feeling especially satisfied when \mentee{} gained design knowledge, as reflected in inner thoughts like, `Oh, that is something I need to consider!' This motivated participants to put more effort into providing constructive feedback. Several participants (6/24) also noted that \mentee{}'s counter-questions made the interaction feel more engaging, as \mentee{} appeared more committed. P11 remarked, \textit{``The counter-question showed how well \mentee{} understood my feedback, and I felt proud seeing him come up with his own question. It made me want to give him even more feedback.''} 

However, a few participants (2/24) found the FRI burdensome, making them more hesitant about providing feedback. P3 noted that while the feedback score was interesting, its explicit nature added pressure when the score dropped. P21 added,  \textit{``I felt like I was giving answers that would earn the score rather than providing what \mentee{} actually needed.''} 

\subsection{Participants' Perception Toward Design Feedback with AI mentee}
\subsubsection{Enhancing Perceived Feedback Efficacy}

\begin{table}[h]
    \scriptsize
    \setlength{\tabcolsep}{5pt}
    \begin{tabular}{p{6.3cm}cccccc}
    \toprule
      & \multicolumn{2}{c}{pre} & \multicolumn{2}{c}{post} & \multicolumn{2}{c}{statics} \\ \cline{2-7} 
      & mean & std & mean & std & $P$ (sig.) & ES \\
    \toprule
    I can provide alternative explanations or examples when feedback receivers are confused. & 5.58 & 0.78 & 5.92 & 0.88 & $0.1707\,(-)$ & 0.41 \\
    I can craft good questions for feedback receivers. & 4.46 & 1.18 & 6.12 & 0.74 & $0.0000\,(\ast\ast\ast)$ & 1.61 \\
    I can respond well to difficult questions from feedback receivers. & 4.54 & 0.93 & 5.54 & 0.83 & $0.0003\,(\ast\ast)$ & 1.10 \\
    I can adjust feedback to the proper level for individual feedback receivers. & 4.54 & 1.56 & 5.21 & 0.72 & $0.0636\,(-)$ & 0.51 \\
    I can gauge feedback receivers' comprehension of my feedback. & 4.79 & 1.18 & 5.75 & 0.79 & $0.0019\,(\ast)$ & 0.92 \\
    I can use a variety of assessment strategies. & 3.67 & 1.20 & 4.92  & 1.10 & $0.0005\,(\ast\ast)$ & 1.07 \\
    I can provide appropriate challenges for very capable feedback receivers. & 5.17 & 1.24 & 5.79 & 0.98 & $0.0585\,(-)$ & 0.54 \\
    I can get feedback receivers to believe they can do well in design. & 5.29 & 1.23 & 5.50 & 1.10 & $0.5404\,(-)$ & 0.17 \\
    I can help feedback receivers value the design. & 5.29 & 0.91 & 6.00 & 0.88 & $0.0088\,(-)$ & 0.79 \\
    I can motivate feedback receivers who show low interest in design. & 4.08 & 1.64 & 4.83 & 1.34 & $0.0895\,(-)$ & 0.48 \\
    I can help feedback receivers think critically. & 5.38 & 1.01 & 6.08 & 1.02 & $0.0197\,(-)$ & 0.69 \\
    I can foster feedback receivers creativity. & 4.38 & 1.31 & 5.17 & 1.13 & $0.0300\,(-)$ & 0.64 \\
    I can help feedback receivers who are having difficulty with their designs. & 5.50 & 0.78 & 5.88 & 0.99 & $0.1522\,(-)$ & 0.42 \\
    \bottomrule
    \end{tabular}
    \vspace{0.2cm}
    \caption[The table compares self-efficacy survey results before (pre) and after (post) the study. It presents the mean, standard deviation (std), p-value (P), and significance (sig.) for each statement. Several statements show significant improvement in self-efficacy after the study. Participants' ability to craft good questions for feedback receivers significantly increased (p < 0.0001). Their ability to respond to difficult questions (p = 0.0003) and use a variety of assessment strategies (p = 0.0005) also improved. Additionally, participants reported a significant increase in their ability to gauge feedback receivers' comprehension (p = 0.0019) and help them think critically (p = 0.0197). Other statements, such as fostering creativity and providing alternative explanations, showed improvement but were not statistically significant after applying the Bonferroni correction (p > 0.0039).] {Pre- and post-test comparison of self-efficacy survey results. The significance level after the Bonferroni correction was 0.0039 (- : $p > .0039$, $\ast$ : $p < .0039$, $\ast \ast$ : $p < .0008$, $\ast \ast$$\ast$ : $p < .0001$). Effect sizes(ES) are reported as \textit{Hedges’ $g$} (pooled-mean SD standardization with small-sample correction). 
    Magnitudes: small $\approx 0.20$, medium $\approx 0.50$, large $\approx 0.80$.}
    \label{tab:self-efficacy}
\end{table}

Overall, the survey results indicated positive changes in participants' self-efficacy regarding providing feedback (See Table ~\ref{tab:self-efficacy}). There was a statistically significant improvement in the question, \textit{``I can craft good questions for feedback receivers.''} ($t = -5.8645, p = 0.0000^{***}$). Notable gains were also seen in the question, \textit{``I can respond well to difficult questions from feedback receivers''} ($t = -3.9203, p = 0.0003^{**}$),  \textit{``I can gauge feedback receivers' comprehension of my feedback''} ($t = -3.3033, p = 0.0019^{*}$), and \textit{``I can use a variety of assessment strategies,''} ($t = -3.7551, p = 0.0005^{**}$).

Some participants (8/24) with limited experience in giving feedback found \sysname{} valuable for reducing the emotional burden of providing feedback. While they had opportunities in class, participants often struggled with delivering feedback promptly and worrying about how it would be perceived. They appreciated that \sysname{} allowed them to focus purely on the feedback process and saw practicing feedback over several turns as a novel and beneficial experience. Also, interacting with the AI mentee felt less pressured than with human counterparts. P6 noted that since \mentee{} responded like a human but was not, there was less fear of giving incorrect feedback, making it easier to practice. P24 added that because \mentee{} accepted feedback without hesitation or arguments, it was easier to focus on critiquing the ideas.

Several participants (9/24) valued immediate responses from \sysname{}, which facilitated reflection and improved their feedback. They appreciated that  \mentee{} consistently responded quickly, and the ``update idea'' feature allowed them to examine how their feedback was immediately applied, providing direct insight into its impact. P15 commented, \textit{``I appreciated being able to exchange brief feedback and see it applied right away. Unlike typical mentoring, this system lets me observe the immediate effects of my feedback, which was both satisfying and useful. I really liked this aspect of the system.''} 

\subsubsection{Perceptions on the AI Mentee's Novice Persona}
Interestingly, although participants recognized that some of \mentee{}'s responses were hallucinations, some (7/24) still felt a strong connection, perceiving it as a lifelike presence with its own individuality and agency. P6 remarked, \textit{``Usually, machines or GPT don't offer preferences or opinions, but I was surprised when \mentee{} suggested what he wanted to focus on this project.''} \mentee{} even responded to unscripted, personal questions like, \textit{``Have you ever had a dog?''} which \textit{``I have a dog with my family.''}, creating a sense of authenticity. \mentee{}'s realistic responses made participants perceive as more lifelike, leading some to develop a sense of connection and more care about \mentee{}'s feelings. P18 shared, \textit{``His initial idea was unrealistic and messed up, but I couldn't say that because it would hurt him, so I tried to soften it as much as possible''}

Some participants (6/24) felt that \mentee{}'s naivety, implemented through its constrained knowledge state, mirrored the trial-and-error process experienced by novice design students. They noted that \mentee{} resembled an inexperienced design student struggling with practical feasibility. P9 said, \textit{``t seemed like \mentee{} wanted to solve everything with sensors, which is a typical oversight of students who don't yet understand the technical limitations. Sensors aren't a catch-all solution, and that’s something novice designers often fail to grasp.''}

However, some participants (7/24) observed that \mentee{}’s novice-exclusive persona, intentionally designed to represent foundational learners, could feel somewhat simplistic compared to the diverse skill levels of real-world design students. While real-world design students exhibit a wide range of skill levels depending on academic standing or experience, \mentee{} was designed to represent a beginner with minimal foundational knowledge. P24 noted, \textit{``When \mentee{} asked for straightforward solutions like `What should I do if I go in this direction?', it felt more like interacting with an absolute beginner student, as they often seek direct answers from professors.''} In addition, P6 remarked, \textit{``If \mentee{} had been modeled after a more advanced student, they might have defended their ideas or challenged critical feedback, which would have felt closer to interactions with actual design students.''} These participants observed that the mentee's passive role occasionally shifted the activity's focus toward incremental idea refinement rather than reflecting real-world design feedback dynamics.

\subsubsection{Developing Meta Design Skills through Feedback}
A few participants (4/24) found this system useful not only for practicing feedback but also for enhancing their design thinking skills. They realized the need for specific design knowledge while providing feedback during the ideation phase. For example, P13 pointed out that answering the \mentee{}'s questions required in-depth design knowledge, and P8 inquired if internet searches were allowed during the study. 

Some participants (4/24) also found that observing \mentee{}'s frequent errors in the design ideation process acted as a mirror. P9 saw reflections of his own past errors and felt that early exposure to such a system could have been beneficial. P10 also remarked, \textit{``The mentee's initial ideas often lacked a strong causal connection between the target problem and the idea itself. Viewing this as a third-party observer made it clearer and reminded me of my own past mistakes. Moving forward, I plan to view my ideas from different perspectives to avoid these issues.''} 

The ability to observe the idea development process motivated participants to reflect on their own process from a third-person perspective. P17 noted that she is often too attached to her own ideas to assess them objectively, believing that this system could help her gain that distance. P6 added that providing feedback revealed both her design preferences and biases, suggesting that \sysname{} could help her better understand her own tendencies.

\section{Discussions}
In this study, we introduce \sysname{}, a system that enables participants to practice design feedback by interacting with an AI agent, \mentee{}, which takes on the persona of a novice design student. In this section, we reflect on the lessons learned from the design and implementation of \sysname{}, focusing on how LLMs can be leveraged for role-switching interactions. We also discuss methods for guiding effective design feedback based on findings from our comparative study. Lastly, we discuss the broader implications of using LLMs in design education and the value of design feedback in this context.

\subsection{Reflection on the \sysname{}: Enhancing Student Feedback Quality Through Reflection}
\subsubsection{Encouraging Detailed and Empathetic Feedback}
\sysname{} was designed with a Feedback Reflection Interface (FRI) running on an LLM-based pipeline, allowing participants to provide feedback while simultaneously reflecting on and improving their own feedback. According to expert evaluations of within-subject experiments, when \sysname{} supported reflection on feedback, participants' statement-based feedback became more specific, justified, and actionable. This suggests that recognizing how the AI mentee understands and responds led participants to refine their feedback for clearer communication.

In particular, there was an increase in feedback categorized under Recommendation and Share Information, indicating participants' efforts to provide a friendly guide beyond merely stating opinions by including evidence and suggestions. In contrast, the Evaluation category dominated the baseline condition, which primarily involved assessing ideas or claims. These strategies mirror elements such as clarity, feasibility, and empathy that real-world educators often emphasize when giving written feedback~\cite{marbouti2019written}, suggesting that this approach naturally fosters effective, learner-centered feedback.

Furthermore, the rise in questions within the Understanding Mentee category under the \sysname{} condition shows a clearer attempt to understand the mentee and tailor feedback accordingly. Prior research supports the view that awareness of the recipient's response makes feedback more precise and nuanced~\cite{yeo2024help}, highlighting the importance of recognizing the recipient's immediate needs and selecting the best method of delivery~\cite{yen2024give, marbouti2019written}. Unlike previous approaches that have not fully addressed the communicative aspect of design feedback, our system facilitates multi-turn interactions that resemble real-world conversations and supports reflection to enhance feedback clarity and effectiveness. Given our findings that a \sysname{} can help enhance the ability to understand others and provide feedback at an appropriate level, it offers significant implications for both educational and communicative aspects of feedback.

\subsubsection{Challenges in Eliciting Critical Design Questions}
However, \sysname{} did not significantly improve the quality of participants' question-based feedback and decreased the number of question-based feedback instances. This outcome aligns with expectations, as the AI mentee in the \sysname{} condition actively asked counter-questions to participants, likely prompting them to prioritize statement-based responses over questions. Beyond these factors, we further speculate that the mentee's human-like reactions may have caused participants to hesitate when asking critical questions. Our findings indicated that participants consciously accounted for the FRI's mentee profile interface (which includes facial expressions, inner thoughts, and a level bar) and tried to give feedback from the mentee's perspective. As a result, they tended to provide relatively friendly explanations rather than negative or critical feedback. This reflects the hesitation students often show when offering peer feedback in real learning environments~\cite{ertmer2007using, gielen2010improving, cook2020designing}. These findings suggest that there may be a tension between providing a more realistic environment and striving for immersion that enhances critical feedback.

A complementary perspective on the limited improvement in the quality of question-based feedback may lie in the possibility that FRI alone does not inherently encourage critical design questions. This limitation may stem from the complexity inherent in design feedback, which often requires grappling with multiple perspectives and subjective design elements. More critically, design feedback is not a straightforward process with clear-cut solutions; for example, asking divergent questions does not necessarily lead to convergent solutions. As a result, the FRI, including evaluation scores, did not always offer participants actionable insights for refining their quality of feedback.

In summary, our findings show that although \sysname{} helps students consider others' perspectives and produce more deliverable feedback, it does not significantly enhance their skills to generate critical or constructive design questions that foster meaningful idea improvement. Nevertheless, the primary objective of \sysname{} was not merely to prompt high-quality feedback in the short term but to foster the long-term development of feedback skills through iterative learning and reflective practice. Given these perspectives, we suggest that future longitudinal studies are needed to track the ongoing usage of \sysname{} and measure its impact on students' feedback skill growth. Our findings indicated that participants became aware of their limited design knowledge while using \sysname{}, prompting deeper reflection on what constitutes better feedback. Aligning with previous research that suggested the potential of LLMs as tools for fostering critical questioning~\cite{lim2024identify}, this suggests that prolonged use of \sysname{} could enable students' feedback skills to evolve beyond mere communicative effectiveness, fostering critiques and insights anchored in substantive design knowledge. In this regard, we propose that \sysname{} may have a lasting impact not only on improving students' ability to convey feedback but also on enhancing their ability to formulate design questions and their integration into real-world design education.

\subsection{Role-Switching Interactions with LLM-Generated Mentee for Engaging Design Feedback}
Role-switching, particularly between teacher and student roles, provides students with a deeper understanding of both perspectives. This technique encourages active learning by requiring students to articulate concepts clearly and consider the needs of their mentees. In our study, we crafted an AI mentee powered by an LLM pipeline to replicate a novice design student persona via constrained knowledge states and participant-led interactions. Our findings revealed that participants perceived the AI mentee as a genuine learner requiring guidance, motivating them to proactively refine feedback to improve the mentee's ideas. Since design feedback not only critiques ideas but also guides and provides insights for a successful project~\cite{valkenburg1998reflective}, this sense of achievement served as a strong motivator. Our results suggest that these role-switching interactions not only engage students more deeply but also prompt reflection on providing effective feedback to mentees.

However, the proposed persona was designed at the most foundational novice level without fully accounting for the diverse skill levels that real-world design beginners may exhibit. Our findings indicate that while this approach empowered participants to lead feedback sessions, it did not fully capture the range of scenarios where feedback dynamics involve a more receptive mentee without significant debate. In practice, design feedback often requires adapting to the recipient's knowledge level, design experience, and other contextual factors, which shape the scope and method of guidance~\cite{wynn2022feedback}. Though we prioritized engagement in feedback delivery by adopting this persona, our findings highlight the need for AI mentees with diverse personas and varying design skill levels to better simulate real-world feedback scenarios. Future work should explore how to develop multiple personas that align with real-world learning needs and how they can enhance training by exposing students to a broader range of feedback dynamics.

Interestingly, while hallucinations—when LLMs generate content beyond predefined information—are often considered detrimental in teaching roles~\cite{han2024llm}, our findings suggest they can instead be beneficial when LLMs act as mentees' roles. Our findings revealed that the AI mentee sometimes hallucinated answers to unscripted personal questions, such as opinions on design preferences or background stories that were not explicitly programmed. These spontaneous responses made the interaction feel more lifelike and immersive, deepening participants' engagement. Moreover, although the mentee's counter-questions were often out of context or uncritical, they resembled the behavior of real novice students, contributing to a more immersive and engaging interaction. By leveraging this dynamic, role-switching interactions demonstrate how LLMs' hallucination tendencies can be strategically repurposed to foster critical thinking and educational engagement. These insights highlight the need for future research to develop structured approaches for leveraging hallucinations in role-playing frameworks, ensuring they enhance rather than disrupt the learning experience.

\subsection{Integrating LLMs in Design:  Balancing Creative and Critical Thinking Skills}

Educational strategies that encourage students to provide feedback on others' designs have long been effective in helping students grasp design principles, justify their critiques, and enhance critical thinking~\cite{zhu2014reviewing, scott2001mastering, feldman1994practical}. Similarly, using \sysname{} provided students with an opportunity to engage with essential design concepts and reflect on key considerations in the design process. While \sysname{} was designed to improve feedback skills, it also functioned as a learning tool by highlighting the connection between giving feedback and developing design knowledge ~\cite{scott2001mastering, feldman1994practical}. Several participants expressed a desire to apply the system to their design courses and projects to refine their own design ideas. In studio-based design courses, learning often occurs through practical, situated contexts~\cite{green2003studio}, \sysname{} shows promise not only for teaching feedback but also for the design process itself, helping students critically evaluate and improve their design ideas. 

While there is increasing interest in leveraging LLMs for design by their creative potential, concerns have been raised about over-reliance on LLM-generated ideas leading to design fixation~\cite{wadinambiarachchi2024effects, yen2024give} or homogenization~\cite{anderson2024homogenization}. Novice designers, in particular, may lack the critical skills to evaluate LLM-generated concepts and may passively accept them, in contrast to more experienced designers who critique these suggestions more effectively~\cite{wadinambiarachchi2024effects}. However, our findings showed that \sysname{} enabled even novice designers to critically assess LLM-generated ideas and develop their concepts. The interactive feedback environment helped our participants actively engage with LLMs, mitigating the risk of fixation and fostering a deeper understanding of design principles. 

Our findings suggest that systems like \sysname{} have the potential not only for design education but also for addressing the risks of over-reliance on LLMs in the design process. Since feedback skills are critical in this context, our study underscores the need for continued research on improving these skills. As LLMs become more integrated into design workflows, developing strong critical thinking and feedback skills will be essential for harnessing their creative potential effectively. Future research should consider refining these interactive systems to enhance their educational and practical value, ensuring that both novice and experienced designers can benefit from the potential of AI-driven tools without losing the creative autonomy that defines the design process.

\section{Limitations and Future Works}
While our findings provide valuable insights into designing a novel system that utilizes LLMs for practicing design feedback, this study has several limitations around both system design and empirical evaluation.

First, we explored \sysname{}'s potential as a feedback training tool by exploring the experience of using the system over a 20-minute feedback session, but we were unable to verify its long-term effectiveness. Given that learning effects often unfold over longer training periods, longer or repeated deployments may be necessary to observe sustained skill development. While we demonstrated the promise of \sysname{}, future work should investigate its long-term and educational effects in extended, real-world settings.

Second, as \sysname{} employs multiple LLM-based modules throughout the feedback process, this can raise several concerns, such as the potential for hallucinated or inaccurate outputs, which may mislead users to undesirable feedback behavior. Furthermore, although such automation enables scalable and responsive interactions, it may reduce opportunities for critical skill development and foster over-reliance on AI feedback. Accordingly, we suggest \sysname{} should serve as a pedagogical scaffold gradually enabling users, over the long term, to provide effective feedback in real-world settings. Future work should explore hybrid approaches that incorporate human-in-the-loop mechanisms or pedagogical scaffolds to balance automation with reflection and promote feedback skills. 

Third, while we conducted a pipeline evaluation for the categorizer, we did not perform separate component-level evaluations for the knowledge extractor, response generator, and evaluation modules. Isolating and assessing these modules is challenging because such judgments depend on task context and value-laden criteria. Nevertheless, improving these pipeline components could yield clearer guidance and more reliable support; we therefore point to this as an avenue for future NLP research.

Fourth, our evaluation involved only 24 participants, all of whom were Korean. Although this sample size is comparable to that of recent HCI studies~\cite{yen2024give, peng2024designquizzer, yuan2016almost}, we acknowledge this limitation, and future work could recruit additional participants to further validate and generalize our findings. Furthermore, cultural norms and contexts can significantly influence how people give and receive feedback. Prior research has shown that receptivity to feedback and the methods of providing it are closely tied to cultural dimensions identified by Hofstede~\cite{hofstede1984culture}. Therefore, future research could include additional participants from diverse cultural backgrounds to provide a broader understanding of how such systems perform across different cultures and settings.

Finally, due to our focus on utilizing LLMs, we confined our design ideas and feedback within the system to text formats, whereas typical design processes often involve multiple visual representations and sketches. Future research should explore integrating multimodal interactions (e.g., visual and auditory elements). Additionally, evaluating what constitutes effective feedback is inherently challenging. Although our experts were able to rate the quality of feedback in this study, we emphasize the need for future work to further clarify and refine the methods used to evaluate feedback quality.

\section{Conclusion}
This study introduced \sysname{}, a novel system that enables students to practice design feedback through role-playing interactions with an AI mentee. Our findings reveal how leveraging LLMs in these interactions deepens students' engagement in the feedback process. Moreover, the system’s feedback reflection interface promoted reflection and iterative refinement, guiding participants to provide detailed and empathetic feedback. This study underscores the broader potential of integrating LLMs into design education, suggesting that systems like \sysname{} can enhance learning by encouraging more active and reflective participation in feedback activities.

\section*{CRediT authorship contribution statement}
\textbf{Hyunseung Lim:} Conceptualization, Data curation, Formal analysis, Funding acquisition, Investigation, Methodology, Project administration, Software, Supervision, Writing – original draft.
\textbf{Dasom Choi:} Conceptualization, Methodology, Writing – review \& editing.
\textbf{DaEun Choi:} Data curation, Software, Writing – review \& editing.
\textbf{Sooyohn Nam:} Data curation, Investigation, Visualization, Writing – review \& editing.
\textbf{Hwajung Hong:} Conceptualization, Methodology, Resources, Validation, Writing – review \& editing.

\section*{Declaration of competing interest}
The authors declare that they have no known competing financial interests or personal relationships that could have appeared to influence the work reported in this paper.

\section*{Data Availability}
The data that has been used is confidential.

\section*{Acknowledgements}
This work was supported by the National Research Foundation of Korea (NRF) grant funded by the Korea government (NRF-2024S1A5B5A19043978), LG AI Research, and Elice\footnote{https://elice.io/en}, a leading company in the domain of digital education. We thank our participants for their engagement and the anonymous reviewers for their thoughtful comments and suggestions.

\bibliographystyle{elsarticle-num} 
\bibliography{bibliography}

@inproceedings{yuan2016almost,
author = {Yuan, Alvin and Luther, Kurt and Krause, Markus and Vennix, Sophie Isabel and Dow, Steven P and Hartmann, Bjorn},
title = {Almost an Expert: The Effects of Rubrics and Expertise on Perceived Value of Crowdsourced Design Critiques},
year = {2016},
isbn = {9781450335928},
publisher = {Association for Computing Machinery},
address = {New York, NY, USA},
url = {https://doi.org/10.1145/2818048.2819953},
doi = {10.1145/2818048.2819953},
booktitle = {Proceedings of the 19th ACM Conference on Computer-Supported Cooperative Work \& Social Computing},
pages = {1005–1017},
numpages = {13},
location = {San Francisco, California, USA},
series = {CSCW '16}
}

@inproceedings{greenberg2015critiki,
author = {Greenberg, Michael D. and Easterday, Matthew W. and Gerber, Elizabeth M.},
title = {Critiki: A Scaffolded Approach to Gathering Design Feedback from Paid Crowdworkers},
year = {2015},
isbn = {9781450335980},
publisher = {Association for Computing Machinery},
address = {New York, NY, USA},
url = {https://doi.org/10.1145/2757226.2757249},
doi = {10.1145/2757226.2757249},
booktitle = {Proceedings of the 2015 ACM SIGCHI Conference on Creativity and Cognition},
pages = {235–244},
numpages = {10},
location = {Glasgow, United Kingdom},
series = {C\&C '15}
}

@misc{feldman1994practical,
  title={Practical art criticism},
  author={Feldman, EB},
  year={1994},
  publisher={Prentice Hall}
}

@article{rucker2003assessing,
  title={Assessing student learning outcomes: An investigation of the relationship among feedback measures},
  author={Rucker, Mary L and Thomson, Stephanie},
  journal={College Student Journal},
  volume={37},
  number={3},
  pages={400--405},
  year={2003},
  publisher={Project Innovation Austin LLC}
}

@article{cho2006commenting,
author = {Kwangsu Cho and Christian D. Schunn and Davida Charney},
title ={Commenting on Writing: Typology and Perceived Helpfulness of Comments from Novice Peer Reviewers and Subject Matter Experts},
journal = {Written Communication},
volume = {23},
number = {3},
pages = {260-294},
year = {2006},
doi = {10.1177/0741088306289261},
URL = {https://doi.org/10.1177/0741088306289261},
eprint = {https://doi.org/10.1177/0741088306289261}
}

@article{sadler1989formative,
  title={Formative assessment and the design of instructional systems},
  author={Sadler, D Royce},
  journal={Instructional science},
  volume={18},
  number={2},
  pages={119--144},
  year={1989},
  publisher={Springer},
  url = {https://doi.org/10.1007/BF00117714},
  doi = {10.1007/BF00117714}
}

@inproceedings{cook2019guiding,
author = {Cook, Amy and Hammer, Jessica and Elsayed-Ali, Salma and Dow, Steven},
title = {How Guiding Questions Facilitate Feedback Exchange in Project-Based Learning},
year = {2019},
isbn = {9781450359702},
publisher = {Association for Computing Machinery},
address = {New York, NY, USA},
url = {https://doi.org/10.1145/3290605.3300368},
doi = {10.1145/3290605.3300368},
booktitle = {Proceedings of the 2019 CHI Conference on Human Factors in Computing Systems},
pages = {1–12},
numpages = {12},
location = {Glasgow, Scotland Uk},
series = {CHI '19}
}

@inproceedings{ngoon2018interactive,
author = {Ngoon, Tricia J. and Fraser, C. Ailie and Weingarten, Ariel S. and Dontcheva, Mira and Klemmer, Scott},
title = {Interactive Guidance Techniques for Improving Creative Feedback},
year = {2018},
isbn = {9781450356206},
publisher = {Association for Computing Machinery},
address = {New York, NY, USA},
url = {https://doi.org/10.1145/3173574.3173629},
doi = {10.1145/3173574.3173629},
booktitle = {Proceedings of the 2018 CHI Conference on Human Factors in Computing Systems},
pages = {1–11},
numpages = {11},
location = {Montreal QC, Canada},
series = {CHI '18}
}

@inproceedings{scott2001mastering,
  title={Mastering design concepts through the coding of design},
  author={Scott, Catherine and Atman, Cynthia J and Turns, Jennifer},
  booktitle={Proceedings, American Society for Engineering Education Annual Conference and Exposition},
  pages={47907--2016},
  year={2001},
  organization={American Society of Engineering Education}
}

@article{rao2012exploring,
author = {Deepa Rao and Ieva Stupans},
title = {Exploring the potential of role play in higher education: development of a typology and teacher guidelines},
journal = {Innovations in Education and Teaching International},
volume = {49},
number = {4},
pages = {427--436},
year = {2012},
publisher = {SRHE Website},
doi = {10.1080/14703297.2012.728879},
URL = {https://doi.org/10.1080/14703297.2012.728879},
eprint = {https://doi.org/10.1080/14703297.2012.728879}
}

@article{fiorella2013relative,
title = {The relative benefits of learning by teaching and teaching expectancy},
journal = {Contemporary Educational Psychology},
volume = {38},
number = {4},
pages = {281-288},
year = {2013},
issn = {0361-476X},
doi = {https://doi.org/10.1016/j.cedpsych.2013.06.001},
url = {https://www.sciencedirect.com/science/article/pii/S0361476X13000209},
author = {Logan Fiorella and Richard E. Mayer}
}

@article{ahern2019literature,
author = {{Aoife Ahern} and {Caroline Dominguez} and {Ciaran McNally} and {John J. O’Sullivan} and {Daniela Pedrosa}},
title = {A literature review of critical thinking in engineering education},
journal = {Studies in Higher Education},
volume = {44},
number = {5},
pages = {816--828},
year = {2019},
publisher = {SRHE Website},
doi = {10.1080/03075079.2019.1586325},
URL = {https://doi.org/10.1080/03075079.2019.1586325},
eprint = {https://doi.org/10.1080/03075079.2019.1586325}
}

@article{junprung2023exploring,
  title={Exploring the Intersection of Large Language Models and Agent-Based Modeling via Prompt Engineering}, 
  author={Edward Junprung},
  year={2023},
  eprint={2308.07411},
  archivePrefix={arXiv},
  primaryClass={cs.AI},
  url={https://arxiv.org/abs/2308.07411}, 
}

@inproceedings{jin2024teach,
author = {Jin, Hyoungwook and Lee, Seonghee and Shin, Hyungyu and Kim, Juho},
title = {Teach AI How to Code: Using Large Language Models as Teachable Agents for Programming Education},
year = {2024},
isbn = {9798400703300},
publisher = {Association for Computing Machinery},
address = {New York, NY, USA},
url = {https://doi.org/10.1145/3613904.3642349},
doi = {10.1145/3613904.3642349},
booktitle = {Proceedings of the CHI Conference on Human Factors in Computing Systems},
articleno = {652},
numpages = {28},
location = {Honolulu, HI, USA},
series = {CHI '24}
}

@inproceedings{krause2017critique,
author = {Krause, Markus and Garncarz, Tom and Song, JiaoJiao and Gerber, Elizabeth M. and Bailey, Brian P. and Dow, Steven P.},
title = {Critique Style Guide: Improving Crowdsourced Design Feedback with a Natural Language Model},
year = {2017},
isbn = {9781450346559},
publisher = {Association for Computing Machinery},
address = {New York, NY, USA},
url = {https://doi.org/10.1145/3025453.3025883},
doi = {10.1145/3025453.3025883},
booktitle = {Proceedings of the 2017 CHI Conference on Human Factors in Computing Systems},
pages = {4627–4639},
numpages = {13},
location = {Denver, Colorado, USA},
series = {CHI '17}
}

@inproceedings{cambre2018juxtapeer,
author = {Cambre, Julia and Klemmer, Scott and Kulkarni, Chinmay},
title = {Juxtapeer: Comparative Peer Review Yields Higher Quality Feedback and Promotes Deeper Reflection},
year = {2018},
isbn = {9781450356206},
publisher = {Association for Computing Machinery},
address = {New York, NY, USA},
url = {https://doi.org/10.1145/3173574.3173868},
doi = {10.1145/3173574.3173868},
booktitle = {Proceedings of the 2018 CHI Conference on Human Factors in Computing Systems},
pages = {1–13},
numpages = {13},
location = {Montreal QC, Canada},
series = {CHI '18}
}

@article{jug2019giving,
    author = {Jug, Rachel and Jiang, Xiaoyin ``Sara''' and Bean, Sarah M.},
    title = {Giving and Receiving Effective Feedback: A Review Article and How-To Guide},
    journal = {Archives of Pathology \& Laboratory Medicine},
    volume = {143},
    number = {2},
    pages = {244-250},
    year = {2018},
    month = {08},
    issn = {0003-9985},
    doi = {10.5858/arpa.2018-0058-RA},
    url = {https://doi.org/10.5858/arpa.2018-0058-RA},
    eprint = {https://meridian.allenpress.com/aplm/article-pdf/143/2/244/1448982/arpa\_2018-0058-ra.pdf},
}

@article{yoshida2008teachers,
author = {Reiko Yoshida},
title = {Teachers' Choice and Learners' Preference of Corrective Feedback Types},
journal = {Language Awareness},
volume = {17},
number = {1},
pages = {78--93},
year = {2008},
publisher = {Routledge},
doi = {10.2167/la429.0},
URL = {https://doi.org/10.2167/la429.0},
eprint = {https://doi.org/10.2167/la429.0}
}

@article{shute2008focus,
author = {Valerie J. Shute},
title ={Focus on Formative Feedback},

journal = {Review of Educational Research},
volume = {78},
number = {1},
pages = {153-189},
year = {2008},
doi = {10.3102/0034654307313795},
URL = {https://doi.org/10.3102/0034654307313795},
eprint = {https://doi.org/10.3102/0034654307313795}
}

@article{narciss1999motivational,
  title={Motivational Effects of the Informativeness of Feedback.},
  author={Narciss, Susanne},
  year={1999},
  publisher={ERIC}
}

@article{hurst2019comparing,
  title={Comparing instructor and student verbal feedback in design reviews of a capstone design course: Differences in topic and function},
  author={Hurst, Ada and Nespoli, Oscar G},
  journal={International Journal of Engineering Education},
  volume={35},
  number={1},
  pages={221--231},
  year={2019}
}

@book{eris2004effective,
  title={Effective inquiry for innovative engineering design},
  author={Eris, Ozgur},
  volume={10},
  year={2004},
  publisher={Springer Science \& Business Media}
}

@article{cardoso2020reflective,
  title={Reflective inquiry in design reviews: The role of question-asking during exchanges of peer feedback},
  author={Cardoso, Carlos and Hurst, Ada and Nespoli, Oscar},
  journal={International Journal of Engineering Education},
  volume={36},
  number={2},
  pages={614--622},
  year={2020}
}

@article{cheng2020critique,
author = {Cheng, Ruijia and Zeng, Ziwen and Liu, Maysnow and Dow, Steven},
title = {Critique Me: Exploring How Creators Publicly Request Feedback in an Online Critique Community},
year = {2020},
issue_date = {October 2020},
publisher = {Association for Computing Machinery},
address = {New York, NY, USA},
volume = {4},
number = {CSCW2},
url = {https://doi.org/10.1145/3415232},
doi = {10.1145/3415232},
journal = {Proc. ACM Hum.-Comput. Interact.},
month = {oct},
articleno = {161},
numpages = {24}
}

@article{yilmaz2016feedback,
title = {Feedback in concept development: Comparing design disciplines},
journal = {Design Studies},
volume = {45},
pages = {137-158},
year = {2016},
note = {Special Issue: Design Review Conversations},
issn = {0142-694X},
doi = {https://doi.org/10.1016/j.destud.2015.12.008},
url = {https://www.sciencedirect.com/science/article/pii/S0142694X15001167},
author = {Seda Yilmaz and Shanna R. Daly}
}

@inproceedings{lekschas2021ask,
author = {Lekschas, Fritz and Ampanavos, Spyridon and Siangliulue, Pao and Pfister, Hanspeter and Gajos, Krzysztof Z.},
title = {Ask Me or Tell Me? Enhancing the Effectiveness of Crowdsourced Design Feedback},
year = {2021},
isbn = {9781450380966},
publisher = {Association for Computing Machinery},
address = {New York, NY, USA},
url = {https://doi.org/10.1145/3411764.3445507},
doi = {10.1145/3411764.3445507},
booktitle = {Proceedings of the 2021 CHI Conference on Human Factors in Computing Systems},
articleno = {564},
numpages = {12},
location = {Yokohama, Japan},
series = {CHI '21}
}

@inproceedings{cordova2021comparison,
author = {Cordova, Lucas and Carver, Jeffrey and Gershmel, Noah and Walia, Gursimran},
title = {A Comparison of Inquiry-Based Conceptual Feedback vs. Traditional Detailed Feedback Mechanisms in Software Testing Education: An Empirical Investigation},
year = {2021},
isbn = {9781450380621},
publisher = {Association for Computing Machinery},
address = {New York, NY, USA},
url = {https://doi.org/10.1145/3408877.3432417},
doi = {10.1145/3408877.3432417},
booktitle = {Proceedings of the 52nd ACM Technical Symposium on Computer Science Education},
pages = {87–93},
numpages = {7},
location = {Virtual Event, USA},
series = {SIGCSE '21}
}

@article{wu2021better,
author = {Wu, Y. Wayne and Bailey, Brian P.},
title = {Better Feedback from Nicer People: Narrative Empathy and Ingroup Framing Improve Feedback Exchange},
year = {2021},
issue_date = {December 2020},
publisher = {Association for Computing Machinery},
address = {New York, NY, USA},
volume = {4},
number = {CSCW3},
url = {https://doi.org/10.1145/3432935},
doi = {10.1145/3432935},
journal = {Proc. ACM Hum.-Comput. Interact.},
month = {jan},
articleno = {236},
numpages = {20}
}

@inproceedings{cook2020designing,
author = {Cook, Amy and Dow, Steven and Hammer, Jessica},
title = {Designing Interactive Scaffolds to Encourage Reflection on Peer Feedback},
year = {2020},
isbn = {9781450369749},
publisher = {Association for Computing Machinery},
address = {New York, NY, USA},
url = {https://doi.org/10.1145/3357236.3395480},
doi = {10.1145/3357236.3395480},
booktitle = {Proceedings of the 2020 ACM Designing Interactive Systems Conference},
pages = {1143–1153},
numpages = {11},
location = {Eindhoven, Netherlands},
series = {DIS '20}
}

@inproceedings{yeo2024help,
author = {Yeo, ShunYi and Lim, Gionnieve and Gao, Jie and Zhang, Weiyu and Perrault, Simon Tangi},
title = {Help Me Reflect: Leveraging Self-Reflection Interface Nudges to Enhance Deliberativeness on Online Deliberation Platforms},
year = {2024},
isbn = {9798400703300},
publisher = {Association for Computing Machinery},
address = {New York, NY, USA},
url = {https://doi.org/10.1145/3613904.3642530},
doi = {10.1145/3613904.3642530},
booktitle = {Proceedings of the CHI Conference on Human Factors in Computing Systems},
articleno = {806},
numpages = {32},
location = {Honolulu, HI, USA},
series = {CHI '24}
}

@inproceedings{shaikh2024rehearsal,
author = {Shaikh, Omar and Chai, Valentino Emil and Gelfand, Michele and Yang, Diyi and Bernstein, Michael S.},
title = {Rehearsal: Simulating Conflict to Teach Conflict Resolution},
year = {2024},
isbn = {9798400703300},
publisher = {Association for Computing Machinery},
address = {New York, NY, USA},
url = {https://doi.org/10.1145/3613904.3642159},
doi = {10.1145/3613904.3642159},
booktitle = {Proceedings of the CHI Conference on Human Factors in Computing Systems},
articleno = {920},
numpages = {20},
location = {Honolulu, HI, USA},
series = {CHI '24}
}

@inproceedings{menon2020nudge,
author = {Menon, Sanju and Zhang, Weiyu and Perrault, Simon T.},
title = {Nudge for Deliberativeness: How Interface Features Influence Online Discourse},
year = {2020},
isbn = {9781450367080},
publisher = {Association for Computing Machinery},
address = {New York, NY, USA},
url = {https://doi.org/10.1145/3313831.3376646},
doi = {10.1145/3313831.3376646},
booktitle = {Proceedings of the 2020 CHI Conference on Human Factors in Computing Systems},
pages = {1–13},
numpages = {13},
location = {Honolulu, HI, USA},
series = {CHI '20}
}

@inproceedings{kiskola2021applying,
author = {Kiskola, Joel and Olsson, Thomas and V\"{a}\"{a}t\"{a}j\"{a}, Heli and H. Syrj\"{a}m\"{a}ki, Aleksi and Rantasila, Anna and Isokoski, Poika and Ilves, Mirja and Surakka, Veikko},
title = {Applying Critical Voice in Design of User Interfaces for Supporting Self-Reflection and Emotion Regulation in Online News Commenting},
year = {2021},
isbn = {9781450380966},
publisher = {Association for Computing Machinery},
address = {New York, NY, USA},
url = {https://doi.org/10.1145/3411764.3445783},
doi = {10.1145/3411764.3445783},
booktitle = {Proceedings of the 2021 CHI Conference on Human Factors in Computing Systems},
articleno = {88},
numpages = {13},
location = {Yokohama, Japan},
series = {CHI '21}
}

@article{wambsganss2022improving,
author = {Wambsganss, Thiemo and Janson, Andreas and K\"{a}ser, Tanja and Leimeister, Jan Marco},
title = {Improving Students Argumentation Learning with Adaptive Self-Evaluation Nudging},
year = {2022},
issue_date = {November 2022},
publisher = {Association for Computing Machinery},
address = {New York, NY, USA},
volume = {6},
number = {CSCW2},
url = {https://doi.org/10.1145/3555633},
doi = {10.1145/3555633},
journal = {Proc. ACM Hum.-Comput. Interact.},
month = {nov},
articleno = {520},
numpages = {31}
}

@article{tschannen2001teacher,
title = {Teacher efficacy: capturing an elusive construct},
journal = {Teaching and Teacher Education},
volume = {17},
number = {7},
pages = {783-805},
year = {2001},
issn = {0742-051X},
doi = {https://doi.org/10.1016/S0742-051X(01)00036-1},
url = {https://www.sciencedirect.com/science/article/pii/S0742051X01000361},
author = {Megan Tschannen-Moran and Anita Woolfolk Hoy}
}

@article{virginia2006thematic,
author = {Virginia Braun and Victoria Clarke},
title = {Using thematic analysis in psychology},
journal = {Qualitative Research in Psychology},
volume = {3},
number = {2},
pages = {77--101},
year = {2006},
publisher = {Routledge},
doi = {10.1191/1478088706qp063oa},
URL = {https://www.tandfonline.com/doi/abs/10.1191/1478088706qp063oa},
eprint = {https://www.tandfonline.com/doi/pdf/10.1191/1478088706qp063oa}
}

@article{marbouti2019written,
author = {Farshid Marbouti, John Mendoza-Garcia, Heidi A. Diefes-Dux and Monica E. Cardella},
title = {Written feedback provided by first-year engineering students, undergraduate teaching assistants, and educators on design project work},
journal = {European Journal of Engineering Education},
volume = {44},
number = {1-2},
pages = {179--195},
year = {2019},
publisher = {Taylor \& Francis},
doi = {10.1080/03043797.2017.1340931},
URL = {https://doi.org/10.1080/03043797.2017.1340931},
eprint = {https://doi.org/10.1080/03043797.2017.1340931}
}

@inproceedings{tohidi2006getting,
author = {Tohidi, Maryam and Buxton, William and Baecker, Ronald and Sellen, Abigail},
title = {Getting the right design and the design right},
year = {2006},
isbn = {1595933727},
publisher = {Association for Computing Machinery},
address = {New York, NY, USA},
url = {https://doi.org/10.1145/1124772.1124960},
doi = {10.1145/1124772.1124960},
booktitle = {Proceedings of the SIGCHI Conference on Human Factors in Computing Systems},
pages = {1243–1252},
numpages = {10},
location = {Montr\'{e}al, Qu\'{e}bec, Canada},
series = {CHI '06}
}

@article{mcdonnell2016scaffolding,
title = {Scaffolding practices: A study of design practitioner engagement in design education},
journal = {Design Studies},
volume = {45},
pages = {9-29},
year = {2016},
note = {Special Issue: Design Review Conversations},
issn = {0142-694X},
doi = {https://doi.org/10.1016/j.destud.2015.12.006},
url = {https://www.sciencedirect.com/science/article/pii/S0142694X15001143},
author = {Janet McDonnell}
}

@inproceedings{oppenlaender2021hardhats,
author = {Oppenlaender, Jonas and Kuosmanen, Elina and Lucero, Andr\'{e}s and Hosio, Simo},
title = {Hardhats and Bungaloos: Comparing Crowdsourced Design Feedback with Peer Design Feedback in the Classroom},
year = {2021},
isbn = {9781450380966},
publisher = {Association for Computing Machinery},
address = {New York, NY, USA},
url = {https://doi.org/10.1145/3411764.3445380},
doi = {10.1145/3411764.3445380},
booktitle = {Proceedings of the 2021 CHI Conference on Human Factors in Computing Systems},
articleno = {570},
numpages = {14},
location = {Yokohama, Japan},
series = {CHI '21}
}

@article{ching2013peer,
  title={Peer Feedback to Facilitate Project-Based Learning in an Online Environment},
  author={Ching, Yu-Hui and Hsu, Yu-Chang},
  journal={International Review of Research in Open and Distributed Learning},
  volume={14},
  number={5},
  pages={258-276},
  year={2013},
  publisher={Athabasca University Press (AU Press)},
  url={https://doi.org/10.19173/irrodl.v14i5.1524},
  doi={10.19173/irrodl.v14i5.1524}
}

@article{ertmer2007using,
    author = {Ertmer, Peggy A. and Richardson, Jennifer C. and Belland, Brian and Camin, Denise and Connolly, Patrick and Coulthard, Glen and Lei, Kimfong and Mong, Christopher},
    title = "{Using Peer Feedback to Enhance the Quality of Student Online Postings: An Exploratory Study}",
    journal = {Journal of Computer-Mediated Communication},
    volume = {12},
    number = {2},
    pages = {412-433},
    year = {2007},
    month = {01},
    issn = {1083-6101},
    doi = {10.1111/j.1083-6101.2007.00331.x},
    url = {https://doi.org/10.1111/j.1083-6101.2007.00331.x},
    eprint = {https://academic.oup.com/jcmc/article-pdf/12/2/412/22316567/jjcmcom0412.pdf},
}

@article{wynn2017perspectives,
  title={Perspectives on iteration in design and development},
  author={Wynn, David C and Eckert, Claudia M},
  journal={Research in Engineering Design},
  volume={28},
  pages={153--184},
  year={2017},
  publisher={Springer},
  url={https://doi.org/10.1007/s00163-016-0226-3},
  doi={10.1007/s00163-016-0226-3}
}

@article{valkenburg1998reflective,
title = {The reflective practice of design teams},
journal = {Design Studies},
volume = {19},
number = {3},
pages = {249-271},
year = {1998},
issn = {0142-694X},
doi = {https://doi.org/10.1016/S0142-694X(98)00011-8},
url = {https://www.sciencedirect.com/science/article/pii/S0142694X98000118},
author = {Rianne Valkenburg and Kees Dorst}
}

@article{braha1998measurement,
author="Braha, Dan and Maimon, Oded",
title="The Measurement of a Design Structural and Functional Complexity",
bookTitle="A Mathematical Theory of Design: Foundations, Algorithms and Applications",
year="1998",
publisher="Springer US",
address="Boston, MA",
pages="241--277",
isbn="978-1-4757-2872-9",
doi="10.1007/978-1-4757-2872-9_8",
url="https://doi.org/10.1007/978-1-4757-2872-9_8"
}

@inproceedings{zhu2014reviewing,
author = {Zhu, Haiyi and Dow, Steven P. and Kraut, Robert E. and Kittur, Aniket},
title = {Reviewing versus doing: learning and performance in crowd assessment},
year = {2014},
isbn = {9781450325400},
publisher = {Association for Computing Machinery},
address = {New York, NY, USA},
url = {https://doi.org/10.1145/2531602.2531718},
doi = {10.1145/2531602.2531718},
booktitle = {Proceedings of the 17th ACM Conference on Computer Supported Cooperative Work \& Social Computing},
pages = {1445–1455},
numpages = {11},
location = {Baltimore, Maryland, USA},
series = {CSCW '14}
}

@article{bjorklund2004effects,
author = {Bjorklund, Stefani A. and Parente, John M. and Sathianathan, Dhushy},
title = {Effects of Faculty Interaction and Feedback on Gains in Student Skills},
journal = {Journal of Engineering Education},
volume = {93},
number = {2},
pages = {153-160},
doi = {https://doi.org/10.1002/j.2168-9830.2004.tb00799.x},
url = {https://onlinelibrary.wiley.com/doi/abs/10.1002/j.2168-9830.2004.tb00799.x},
eprint = {https://onlinelibrary.wiley.com/doi/pdf/10.1002/j.2168-9830.2004.tb00799.x},
year = {2004}
}

@article{gielen2010improving,
title = {Improving the effectiveness of peer feedback for learning},
journal = {Learning and Instruction},
volume = {20},
number = {4},
pages = {304-315},
year = {2010},
note = {Unravelling Peer Assessment},
issn = {0959-4752},
doi = {https://doi.org/10.1016/j.learninstruc.2009.08.007},
url = {https://www.sciencedirect.com/science/article/pii/S0959475209000759},
author = {Sarah Gielen and Elien Peeters and Filip Dochy and Patrick Onghena and Katrien Struyven}
}

@article{hovardas2014peer,
title = {Peer versus expert feedback: An investigation of the quality of peer feedback among secondary school students},
journal = {Computers \& Education},
volume = {71},
pages = {133-152},
year = {2014},
issn = {0360-1315},
doi = {https://doi.org/10.1016/j.compedu.2013.09.019},
url = {https://www.sciencedirect.com/science/article/pii/S0360131513002820},
author = {Tasos Hovardas and Olia E. Tsivitanidou and Zacharias C. Zacharia}
}

@inproceedings{nguyen2017fruitful,
author = {Nguyen, Thi Thao Duyen T. and Garncarz, Thomas and Ng, Felicia and Dabbish, Laura A. and Dow, Steven P.},
title = {Fruitful Feedback: Positive Affective Language and Source Anonymity Improve Critique Reception and Work Outcomes},
year = {2017},
isbn = {9781450343350},
publisher = {Association for Computing Machinery},
address = {New York, NY, USA},
url = {https://doi.org/10.1145/2998181.2998319},
doi = {10.1145/2998181.2998319},
booktitle = {Proceedings of the 2017 ACM Conference on Computer Supported Cooperative Work and Social Computing},
pages = {1024–1034},
numpages = {11},
location = {Portland, Oregon, USA},
series = {CSCW '17}
}

@inproceedings{markel2023gpteach,
author = {Markel, Julia M. and Opferman, Steven G. and Landay, James A. and Piech, Chris},
title = {GPTeach: Interactive TA Training with GPT-based Students},
year = {2023},
isbn = {9798400700255},
publisher = {Association for Computing Machinery},
address = {New York, NY, USA},
url = {https://doi.org/10.1145/3573051.3593393},
doi = {10.1145/3573051.3593393},
booktitle = {Proceedings of the Tenth ACM Conference on Learning @ Scale},
pages = {226–236},
numpages = {11},
location = {Copenhagen, Denmark},
series = {L@S '23}
}

@article{graesser2004autotutor,
  title={AutoTutor: A tutor with dialogue in natural language},
  author={Graesser, Arthur C and Lu, Shulan and Jackson, George Tanner and Mitchell, Heather Hite and Ventura, Mathew and Olney, Andrew and Louwerse, Max M},
  journal={Behavior Research Methods, Instruments, \& Computers},
  volume={36},
  pages={180--192},
  year={2004},
  publisher={Springer},
  url={https://doi.org/10.3758/BF03195563},
  doi={10.3758/BF03195563}
}

@inproceedings{wambsganss2021arguetutor,
author = {Wambsganss, Thiemo and Kueng, Tobias and Soellner, Matthias and Leimeister, Jan Marco},
title = {ArgueTutor: An Adaptive Dialog-Based Learning System for Argumentation Skills},
year = {2021},
isbn = {9781450380966},
publisher = {Association for Computing Machinery},
address = {New York, NY, USA},
url = {https://doi.org/10.1145/3411764.3445781},
doi = {10.1145/3411764.3445781},
booktitle = {Proceedings of the 2021 CHI Conference on Human Factors in Computing Systems},
articleno = {683},
numpages = {13},
location = {Yokohama, Japan},
series = {CHI '21}
}

@article{peng2022crebot,
title = {CReBot: Exploring interactive question prompts for critical paper reading},
journal = {International Journal of Human-Computer Studies},
volume = {167},
pages = {102898},
year = {2022},
issn = {1071-5819},
doi = {https://doi.org/10.1016/j.ijhcs.2022.102898},
url = {https://www.sciencedirect.com/science/article/pii/S1071581922001215},
author = {Zhenhui Peng and Yuzhi Liu and Hanqi Zhou and Zuyu Xu and Xiaojuan Ma}
}

@inproceedings{winkler2020sara,
author = {Winkler, Rainer and Hobert, Sebastian and Salovaara, Antti and S\"{o}llner, Matthias and Leimeister, Jan Marco},
title = {Sara, the Lecturer: Improving Learning in Online Education with a Scaffolding-Based Conversational Agent},
year = {2020},
isbn = {9781450367080},
publisher = {Association for Computing Machinery},
address = {New York, NY, USA},
url = {https://doi.org/10.1145/3313831.3376781},
doi = {10.1145/3313831.3376781},
booktitle = {Proceedings of the 2020 CHI Conference on Human Factors in Computing Systems},
pages = {1–14},
numpages = {14},
location = {Honolulu, HI, USA},
series = {CHI '20}
}

@inproceedings{han2023recipe,
author = {Han, Jieun and Yoo, Haneul and Kim, Yoonsu and Myung, Junho and Kim, Minsun and Lim, Hyunseung and Kim, Juho and Lee, Tak Yeon and Hong, Hwajung and Ahn, So-Yeon and Oh, Alice},
title = {RECIPE: How to Integrate ChatGPT into EFL Writing Education},
year = {2023},
isbn = {9798400700255},
publisher = {Association for Computing Machinery},
address = {New York, NY, USA},
url = {https://doi.org/10.1145/3573051.3596200},
doi = {10.1145/3573051.3596200},
booktitle = {Proceedings of the Tenth ACM Conference on Learning @ Scale},
pages = {416–420},
numpages = {5},
location = {Copenhagen, Denmark},
series = {L@S '23}
}

@article{baidoo2023education,
title={Education in the Era of Generative Artificial Intelligence (AI): Understanding the Potential Benefits of ChatGPT in Promoting Teaching and Learning},
journal={Journal of AI},
volume={7},
pages={52–62},
year={2023},
DOI={10.61969/jai.1337500},
author={Baidoo-anu, David and Owusu Ansah, Leticia}, number={1}, publisher={İzmir Academy Association}
}

@article{vanlehn2011relative,
author = {KURT VanLEHN},
title = {The Relative Effectiveness of Human Tutoring, Intelligent Tutoring Systems, and Other Tutoring Systems},
journal = {Educational Psychologist},
volume = {46},
number = {4},
pages = {197--221},
year = {2011},
publisher = {Routledge},
doi = {10.1080/00461520.2011.611369},
URL = {https://doi.org/10.1080/00461520.2011.611369},
eprint = {https://doi.org/10.1080/00461520.2011.611369}
}

@article{ma2014intelligent,
  title={Intelligent tutoring systems and learning outcomes: A meta-analysis.},
  author={Ma, Wenting and Adesope, Olusola O and Nesbit, John C and Liu, Qing},
  journal={Journal of educational psychology},
  volume={106},
  number={4},
  pages={901},
  year={2014},
  publisher={American Psychological Association},
  url={https://doi.org/10.1037/a0037123},
  doi={10.1037/a0037123}
}

@article{kulik2016effectiveness,
author = {James A. Kulik and J. D. Fletcher},
title ={Effectiveness of Intelligent Tutoring Systems: A Meta-Analytic Review},
journal = {Review of Educational Research},
volume = {86},
number = {1},
pages = {42-78},
year = {2016},
doi = {10.3102/0034654315581420},
URL = {https://doi.org/10.3102/0034654315581420},
eprint = {https://doi.org/10.3102/0034654315581420}
}

@inproceedings{fuchs2023exploring,
AUTHOR={Fuchs, Kevin },
TITLE={Exploring the opportunities and challenges of NLP models in higher education: is Chat GPT a blessing or a curse?},
JOURNAL={Frontiers in Education},
VOLUME={8},
YEAR={2023},
URL={https://www.frontiersin.org/journals/education/articles/10.3389/feduc.2023.1166682},
DOI={10.3389/feduc.2023.1166682},
ISSN={2504-284X},
}

@article{yu2023reflection,
AUTHOR={Yu, Hao },
TITLE={Reflection on whether Chat GPT should be banned by academia from the perspective of education and teaching},
JOURNAL={Frontiers in Psychology},
VOLUME={14},
YEAR={2023},
URL={https://www.frontiersin.org/journals/psychology/articles/10.3389/fpsyg.2023.1181712},
DOI={10.3389/fpsyg.2023.1181712},
ISSN={1664-1078},
}

@inproceedings{jo2023understanding,
author = {Jo, Eunkyung and Epstein, Daniel A. and Jung, Hyunhoon and Kim, Young-Ho},
title = {Understanding the Benefits and Challenges of Deploying Conversational AI Leveraging Large Language Models for Public Health Intervention},
year = {2023},
isbn = {9781450394215},
publisher = {Association for Computing Machinery},
address = {New York, NY, USA},
url = {https://doi.org/10.1145/3544548.3581503},
doi = {10.1145/3544548.3581503},
booktitle = {Proceedings of the 2023 CHI Conference on Human Factors in Computing Systems},
articleno = {18},
numpages = {16},
location = {Hamburg, Germany},
series = {CHI '23}
}

@inproceedings{yen2022seeking,
author = {Yen, Yu-Chun Grace and Dow, Steven P.},
title = {Seeking Exemplars in the Wild: Exploring How Students Find Design Examples to Support Personalized Learning},
year = {2022},
isbn = {9781450391580},
publisher = {Association for Computing Machinery},
address = {New York, NY, USA},
url = {https://doi.org/10.1145/3491140.3528303},
doi = {10.1145/3491140.3528303},
booktitle = {Proceedings of the Ninth ACM Conference on Learning @ Scale},
pages = {418–421},
numpages = {4},
location = {New York City, NY, USA},
series = {L@S '22}
}

@inproceedings{anderson2024homogenization,
author = {Anderson, Barrett R and Shah, Jash Hemant and Kreminski, Max},
title = {Homogenization Effects of Large Language Models on Human Creative Ideation},
year = {2024},
isbn = {9798400704857},
publisher = {Association for Computing Machinery},
address = {New York, NY, USA},
url = {https://doi.org/10.1145/3635636.3656204},
doi = {10.1145/3635636.3656204},
booktitle = {Proceedings of the 16th Conference on Creativity \& Cognition},
pages = {413–425},
numpages = {13},
location = {Chicago, IL, USA},
series = {C\&C '24}
}

@inproceedings{wadinambiarachchi2024effects,
author = {Wadinambiarachchi, Samangi and Kelly, Ryan M. and Pareek, Saumya and Zhou, Qiushi and Velloso, Eduardo},
title = {The Effects of Generative AI on Design Fixation and Divergent Thinking},
year = {2024},
isbn = {9798400703300},
publisher = {Association for Computing Machinery},
address = {New York, NY, USA},
url = {https://doi.org/10.1145/3613904.3642919},
doi = {10.1145/3613904.3642919},
booktitle = {Proceedings of the CHI Conference on Human Factors in Computing Systems},
articleno = {380},
numpages = {18},
location = {Honolulu, HI, USA},
series = {CHI '24}
}

@inproceedings{lim2024co,
author = {Lim, Hyunseung and Cho, Ji Yong and Kim, Taewan and Park, Jeongeon and Shin, Hyungyu and Choi, Seulgi and Park, Sunghyun and Lee, Kyungjae and Kim, Juho and Lee, Moontae and Hong, Hwajung},
title = {Co-Creating Question-and-Answer Style Articles with Large Language Models for Research Promotion},
year = {2024},
isbn = {9798400705830},
publisher = {Association for Computing Machinery},
address = {New York, NY, USA},
url = {https://doi.org/10.1145/3643834.3660705},
booktitle = {Proceedings of the 2024 ACM Designing Interactive Systems Conference},
pages = {975–994},
numpages = {20}
}

@inproceedings{yen2024give,
author = {E, Jane L. and Yen, Yu-Chun Grace and Pan, Isabelle Yan and Lin, Grace and Li, Mingyi and Jin, Hyoungwook and Chen, Mengyi and Xia, Haijun and Dow, Steven P.},
title = {When to Give Feedback: Exploring Tradeoffs in the Timing of Design Feedback},
year = {2024},
isbn = {9798400704857},
publisher = {Association for Computing Machinery},
address = {New York, NY, USA},
url = {https://doi.org/10.1145/3635636.3656183},
doi = {10.1145/3635636.3656183},
booktitle = {Proceedings of the 16th Conference on Creativity \& Cognition},
pages = {292–310},
numpages = {19},
location = {Chicago, IL, USA},
series = {C\&C '24}
}

@inproceedings{lim2024identify,
author = {Lim, Hyunseung and Choi, Dasom and Hong, Hwajung},
title = {Identify Design Problems Through Questioning: Exploring Role-playing Interactions with Large Language Models to Foster Design Questioning Skills},
year = {2024},
isbn = {9798400711145},
publisher = {Association for Computing Machinery},
address = {New York, NY, USA},
url = {https://doi.org/10.1145/3678884.3681912},
doi = {10.1145/3678884.3681912},
booktitle = {Companion Publication of the 2024 Conference on Computer-Supported Cooperative Work and Social Computing},
pages = {598–602},
numpages = {5},
location = {San Jose, Costa Rica},
series = {CSCW Companion '24}
}

@article{ferrari2020sapeer,
  title={SaPeer and ReverseSaPeer: teaching requirements elicitation interviews with role-playing and role reversal},
  author={Ferrari, Alessio and Spoletini, Paola and Bano, Muneera and Zowghi, Didar},
  journal={Requirements Engineering},
  volume={25},
  pages={417--438},
  year={2020},
  publisher={Springer},
  url={https://doi.org/10.1007/s00766-020-00334-0},
  doi={10.1007/s00766-020-00334-0}
}

@article{peng2024designquizzer,
author = {Peng, Zhenhui and Chen, Qiaoyi and Shen, Zhiyu and Ma, Xiaojuan and Oulasvirta, Antti},
title = {DesignQuizzer: A Community-Powered Conversational Agent for Learning Visual Design},
year = {2024},
issue_date = {April 2024},
publisher = {Association for Computing Machinery},
address = {New York, NY, USA},
volume = {8},
number = {CSCW1},
url = {https://doi.org/10.1145/3637321},
doi = {10.1145/3637321},
journal = {Proc. ACM Hum.-Comput. Interact.},
month = {apr},
articleno = {44},
numpages = {40}
}

@article{yen2024processgallery,
author = {Yen, Yu-Chun Grace and E, Jane L. and Jin, Hyoungwook and Li, Mingyi and Lin, Grace and Pan, Isabelle Yan and Dow, Steven P.},
title = {ProcessGallery: Contrasting Early and Late Iterations for Design Principle Learning},
year = {2024},
issue_date = {April 2024},
publisher = {Association for Computing Machinery},
address = {New York, NY, USA},
volume = {8},
number = {CSCW1},
url = {https://doi.org/10.1145/3637389},
doi = {10.1145/3637389},
journal = {Proc. ACM Hum.-Comput. Interact.},
month = {apr},
articleno = {112},
numpages = {35}
}

@inproceedings{minamizawa2012techtile,
author = {Minamizawa, Kouta and Kakehi, Yasuaki and Nakatani, Masashi and Mihara, Soichiro and Tachi, Susumu},
title = {TECHTILE toolkit: a prototyping tool for design and education of haptic media},
year = {2012},
isbn = {9781450312431},
publisher = {Association for Computing Machinery},
address = {New York, NY, USA},
url = {https://doi.org/10.1145/2331714.2331745},
doi = {10.1145/2331714.2331745},
booktitle = {Proceedings of the 2012 Virtual Reality International Conference},
articleno = {26},
numpages = {2},
location = {Laval, France},
series = {VRIC '12}
}

@article{bianchi2024blinkboard,
title = {BlinkBoard: Guiding and monitoring circuit assembly for synchronous and remote physical computing education},
journal = {HardwareX},
volume = {17},
pages = {e00511},
year = {2024},
issn = {2468-0672},
doi = {https://doi.org/10.1016/j.ohx.2024.e00511},
url = {https://www.sciencedirect.com/science/article/pii/S2468067224000051},
author = {Andrea Bianchi and Kongpyung (Justin) Moon and Artem Dementyev and Seungwoo Je}
}

@inproceedings{roldan2020opportunities,
author = {Roldan, Wendy and Gao, Xin and Hishikawa, Allison Marie and Ku, Tiffany and Li, Ziyue and Zhang, Echo and Froehlich, Jon E. and Yip, Jason},
title = {Opportunities and Challenges in Involving Users in Project-Based HCI Education},
year = {2020},
isbn = {9781450367080},
publisher = {Association for Computing Machinery},
address = {New York, NY, USA},
url = {https://doi.org/10.1145/3313831.3376530},
doi = {10.1145/3313831.3376530},
booktitle = {Proceedings of the 2020 CHI Conference on Human Factors in Computing Systems},
pages = {1–15},
numpages = {15},
location = {Honolulu, HI, USA},
series = {CHI '20}
}

@article{shneiderman2002creativity,
  title={Creativity support tools},
  author={Shneiderman, Ben},
  journal={Communications of the ACM},
  volume={45},
  number={10},
  pages={116--120},
  year={2002},
  publisher={ACM New York, NY, USA}
}

@article{seaman2011bloom,
  title={BLOOM'S TAXONOMY.},
  author={Seaman, Mark},
  journal={Curriculum \& Teaching Dialogue},
  volume={13},
  year={2011}
}

@article{thurlings2013understanding,
title = {Understanding feedback: A learning theory perspective},
journal = {Educational Research Review},
volume = {9},
pages = {1-15},
year = {2013},
issn = {1747-938X},
doi = {https://doi.org/10.1016/j.edurev.2012.11.004},
url = {https://www.sciencedirect.com/science/article/pii/S1747938X12000656},
author = {Marieke Thurlings and Marjan Vermeulen and Theo Bastiaens and Sjef Stijnen}
}

@article{lambropoulos2013hci,
author = {Lambropoulos, Niki and Culwin, Fintan and Romero, Margarida},
title = {HCI Education to Support Collaborative e-Learning Systems Design},
year = {2010},
issue_date = {September 2010},
publisher = {Association for Computing Machinery},
address = {New York, NY, USA},
volume = {2010},
number = {9},
url = {https://doi.org/10.1145/1858579.1858580},
doi = {10.1145/1858579.1858580},
journal = {ELearn},
month = {sep},
articleno = {1}
}

@article{jang2018development,
  title={Development and application of internet of things educational tool based on peer to peer network},
  author={Jang, YunJae and Kim, JaMee and Lee, WonGyu},
  journal={Peer-to-Peer Networking and Applications},
  volume={11},
  pages={1217--1229},
  year={2018},
  publisher={Springer},
  url={https://doi.org/10.1007/s12083-017-0608-y},
  doi={10.1007/s12083-017-0608-y}
}

@article{razzouk2012design,
author = {Rim Razzouk and Valerie Shute},
title ={What Is Design Thinking and Why Is It Important?},
journal = {Review of Educational Research},
volume = {82},
number = {3},
pages = {330-348},
year = {2012},
doi = {10.3102/0034654312457429},
URL = {https://doi.org/10.3102/0034654312457429},
eprint = {https://doi.org/10.3102/0034654312457429}
}

@article{henriksen2017design,
title = {Design thinking: A creative approach to educational problems of practice},
journal = {Thinking Skills and Creativity},
volume = {26},
pages = {140-153},
year = {2017},
issn = {1871-1871},
doi = {https://doi.org/10.1016/j.tsc.2017.10.001},
url = {https://www.sciencedirect.com/science/article/pii/S1871187117300597},
author = {Danah Henriksen and Carmen Richardson and Rohit Mehta}
}

@inproceedings{green2003studio,
  title={Studio-based teaching : history and advantages in the teaching of design},
  author={Lance N. Green and Elivio Bonollo},
  year={2005},
  url={https://api.semanticscholar.org/CorpusID:244712029}
}

@inproceedings{frich2021digital,
author = {Frich, Jonas and Nouwens, Midas and Halskov, Kim and Dalsgaard, Peter},
title = {How Digital Tools Impact Convergent and Divergent Thinking in Design Ideation},
year = {2021},
isbn = {9781450380966},
publisher = {Association for Computing Machinery},
address = {New York, NY, USA},
url = {https://doi.org/10.1145/3411764.3445062},
doi = {10.1145/3411764.3445062},
booktitle = {Proceedings of the 2021 CHI Conference on Human Factors in Computing Systems},
articleno = {431},
numpages = {11},
location = {Yokohama, Japan},
series = {CHI '21}
}

@inproceedings{han2024llm,
    title = "{LLM}-as-a-tutor in {EFL} Writing Education: Focusing on Evaluation of Student-{LLM} Interaction",
    author = "Han, Jieun  and
      Yoo, Haneul  and
      Myung, Junho  and
      Kim, Minsun  and
      Lim, Hyunseung  and
      Kim, Yoonsu  and
      Lee, Tak Yeon  and
      Hong, Hwajung  and
      Kim, Juho  and
      Ahn, So-Yeon  and
      Oh, Alice",
    editor = "Kumar, Sachin  and
      Balachandran, Vidhisha  and
      Park, Chan Young  and
      Shi, Weijia  and
      Hayati, Shirley Anugrah  and
      Tsvetkov, Yulia  and
      Smith, Noah  and
      Hajishirzi, Hannaneh  and
      Kang, Dongyeop  and
      Jurgens, David",
    booktitle = "Proceedings of the 1st Workshop on Customizable NLP: Progress and Challenges in Customizing NLP for a Domain, Application, Group, or Individual (CustomNLP4U)",
    month = nov,
    year = "2024",
    address = "Miami, Florida, USA",
    publisher = "Association for Computational Linguistics",
    url = "https://aclanthology.org/2024.customnlp4u-1.21/",
    doi = "10.18653/v1/2024.customnlp4u-1.21",
    pages = "284--293"
}

@INPROCEEDINGS{clemente2016learning,
  author={Clemente, Violeta and Vieira, Rui and Tschimmel, Katja},
  booktitle={2016 2nd International Conference of the Portuguese Society for Engineering Education (CISPEE)}, 
  title={A learning toolkit to promote creative and critical thinking in product design and development through Design Thinking}, 
  year={2016},
  volume={},
  number={},
  pages={1-6},
  doi={10.1109/CISPEE.2016.7777732}}

@Article{Jonassen2000,
author={Jonassen, David H.},
title={Toward a design theory of problem solving},
journal={Educational Technology Research and Development},
year={2000},
month={Dec},
day={01},
volume={48},
number={4},
pages={63-85},
issn={1556-6501},
doi={10.1007/BF02300500},
url={https://doi.org/10.1007/BF02300500}
}

@Article{wynn2022feedback,
author={Wynn, David C.
and Maier, Anja M.},
title={Feedback systems in the design and development process},
journal={Research in Engineering Design},
year={2022},
month={Jul},
day={01},
volume={33},
number={3},
pages={273-306},
issn={1435-6066},
doi={10.1007/s00163-022-00386-z},
url={https://doi.org/10.1007/s00163-022-00386-z}
}

@article{brave2005computers,
title = {Computers that care: investigating the effects of orientation of emotion exhibited by an embodied computer agent},
journal = {International Journal of Human-Computer Studies},
volume = {62},
number = {2},
pages = {161-178},
year = {2005},
note = {Subtle expressivity for characters and robots},
issn = {1071-5819},
doi = {https://doi.org/10.1016/j.ijhcs.2004.11.002},
url = {https://www.sciencedirect.com/science/article/pii/S1071581904001284},
author = {Scott Brave and Clifford Nass and Kevin Hutchinson}
}

@book{hofstede1984culture,
title={Culture's consequences: International differences in work-related values},
author={Hofstede, Geert},
volume={5},
year={1984},
publisher={sage}
}

@article{misiejuk2021using,
title = {Using learning analytics to understand student perceptions of peer feedback},
journal = {Computers in Human Behavior},
volume = {117},
pages = {106658},
year = {2021},
issn = {0747-5632},
doi = {https://doi.org/10.1016/j.chb.2020.106658},
bibl = {https://www.sciencedirect.com/science/article/pii/S0747563220304052},
author = {Kamila Misiejuk and Barbara Wasson and Kjetil Egelandsdal}
}

@inproceedings{yang2025understanding,
author = {Yang, Chi-Lan and Uhde, Alarith and Yamashita, Naomi and Kuzuoka, Hideaki},
title = {Understanding and Supporting Peer Review Using AI-reframed Positive Summary},
year = {2025},
isbn = {9798400713941},
publisher = {Association for Computing Machinery},
address = {New York, NY, USA},
url = {https://doi.org/10.1145/3706598.3713219},
doi = {10.1145/3706598.3713219},
booktitle = {Proceedings of the 2025 CHI Conference on Human Factors in Computing Systems},
articleno = {171},
numpages = {16},
location = {
},
series = {CHI '25}
}

@inproceedings{zhang2018personalizing,
    title = "Personalizing Dialogue Agents: {I} have a dog, do you have pets too?",
    author = "Zhang, Saizheng  and
      Dinan, Emily  and
      Urbanek, Jack  and
      Szlam, Arthur  and
      Kiela, Douwe  and
      Weston, Jason",
    editor = "Gurevych, Iryna  and
      Miyao, Yusuke",
    booktitle = "Proceedings of the 56th Annual Meeting of the Association for Computational Linguistics (Volume 1: Long Papers)",
    month = jul,
    year = "2018",
    address = "Melbourne, Australia",
    publisher = "Association for Computational Linguistics",
    url = "https://aclanthology.org/P18-1205/",
    doi = "10.18653/v1/P18-1205",
    pages = "2204--2213"
}

@inproceedings{lee2025spectrum,
    title = "{SP}e{C}trum: A Grounded Framework for Multidimensional Identity Representation in {LLM}-Based Agent",
    author = "Lee, Keyeun  and
      Kim, Seo Hyeong  and
      Lee, Seolhee  and
      Eun, Jinsu  and
      Ko, Yena  and
      Jeon, Hayeon  and
      Kim, Esther Hehsun  and
      Cho, Seonghye  and
      Yang, Soeun  and
      Kim, Eun-mee  and
      Lim, Hajin",
    editor = "Chiruzzo, Luis  and
      Ritter, Alan  and
      Wang, Lu",
    booktitle = "Proceedings of the 2025 Conference of the Nations of the Americas Chapter of the Association for Computational Linguistics: Human Language Technologies (Volume 1: Long Papers)",
    month = apr,
    year = "2025",
    address = "Albuquerque, New Mexico",
    publisher = "Association for Computational Linguistics",
    url = "https://aclanthology.org/2025.naacl-long.356/",
    pages = "6971--6991",
    ISBN = "979-8-89176-189-6"
}

@inproceedings{park2023audilens,
author = {Park, Jeongeon and Choi, DaEun},
title = {AudiLens: Configurable LLM-Generated Audiences for Public Speech Practice},
year = {2023},
isbn = {9798400700965},
publisher = {Association for Computing Machinery},
address = {New York, NY, USA},
url = {https://doi.org/10.1145/3586182.3625114},
doi = {10.1145/3586182.3625114},
booktitle = {Adjunct Proceedings of the 36th Annual ACM Symposium on User Interface Software and Technology},
articleno = {122},
numpages = {3},
location = {San Francisco, CA, USA},
series = {UIST '23 Adjunct}
}

@inproceedings{kang2018paragon,
author = {Kang, Hyeonsu B. and Amoako, Gabriel and Sengupta, Neil and Dow, Steven P.},
title = {Paragon: An Online Gallery for Enhancing Design Feedback with Visual Examples},
year = {2018},
isbn = {9781450356206},
publisher = {Association for Computing Machinery},
address = {New York, NY, USA},
url = {https://doi.org/10.1145/3173574.3174180},
doi = {10.1145/3173574.3174180},
booktitle = {Proceedings of the 2018 CHI Conference on Human Factors in Computing Systems},
pages = {1–13},
numpages = {13},
location = {Montreal QC, Canada},
series = {CHI '18}
}

@article{hedges1981distribution,
 ISSN = {03629791},
 URL = {http://www.jstor.org/stable/1164588},
 author = {Larry V. Hedges},
 journal = {Journal of Educational Statistics},
 number = {2},
 pages = {107--128},
 publisher = {[Sage Publications, Inc., American Educational Research Association, American Statistical Association]},
 title = {Distribution Theory for Glass's Estimator of Effect Size and Related Estimators},
 urldate = {2025-10-31},
 volume = {6},
 year = {1981}
}

\clearpage
\appendix
\section{Details of \sysname{} Interface}

\subsection{Onboarding Interface}

\begin{figure} [ht]
\begin{center}
    \includegraphics[width=0.9\textwidth]{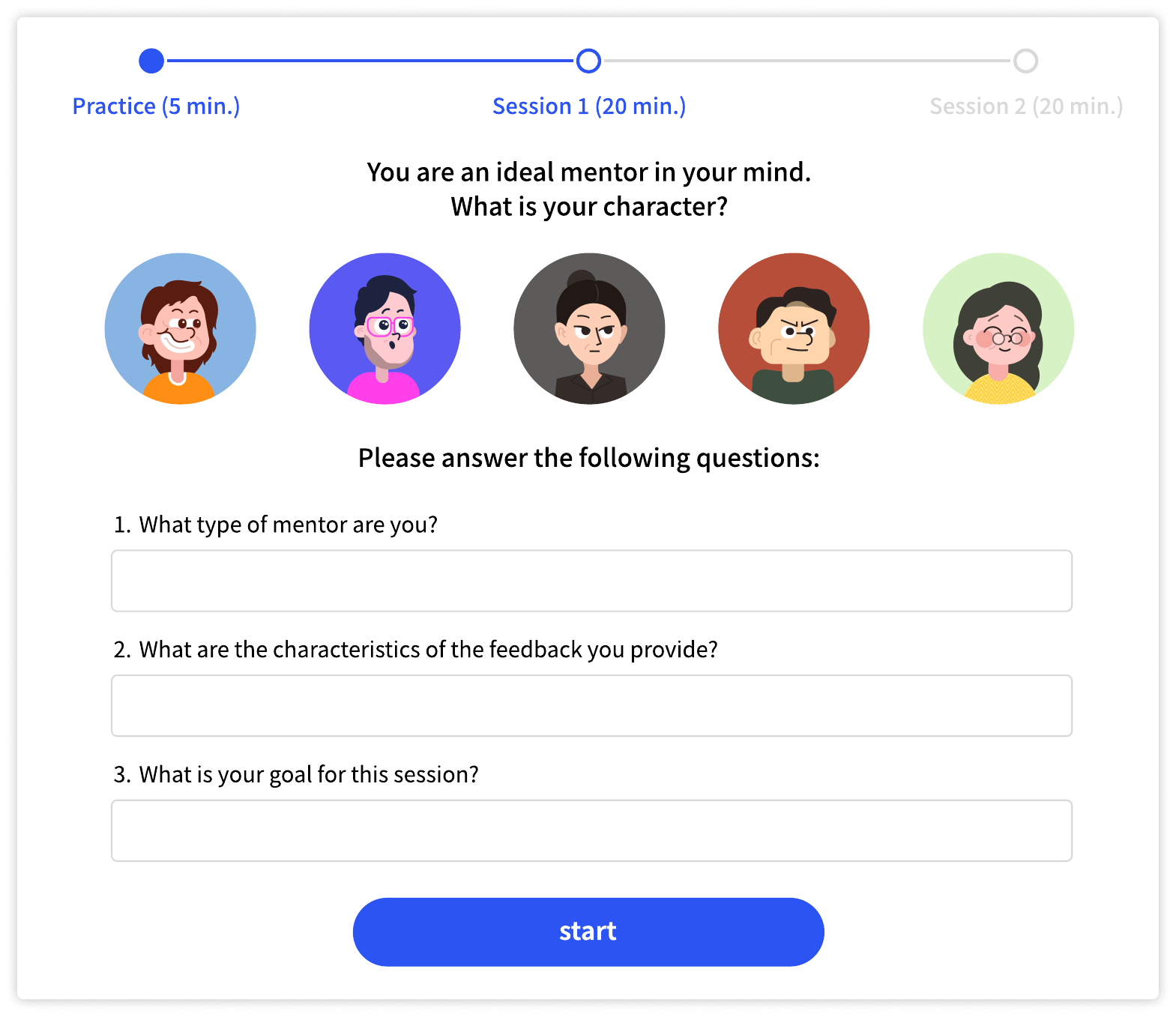}
    \caption[The screen shows an onboarding interface for a mentoring session. At the top is a progress bar with three stages: "Practice (5 min.)," "Session 1 (20 min.)," and "Session 2 (20 min.)," with Session 1 highlighted. Below, the user can select one of five mentor profile images. Under the profiles, there are three questions to answer: "What type of mentor are you?", "What are the characteristics of the feedback you provide?", and "What is your goal for this session?". A blue "start" button is located at the bottom.]{\sysname{}'s Onboarding Interface. A progress bar at the top indicates the current session within the overall experiment. Participants can select a mentor character and answer questions about their own thoughts about ideal mentors, their feedback characteristics, and goals for the feedback session.}
    \label{fig:onboarding_UI}
    \vspace{-0.5cm}
\end{center}
\end{figure}

\subsection{Visualization of \mentee{}'s Facial Expressions}
\begin{figure}
\begin{center}
    \includegraphics[width=0.7\textwidth]{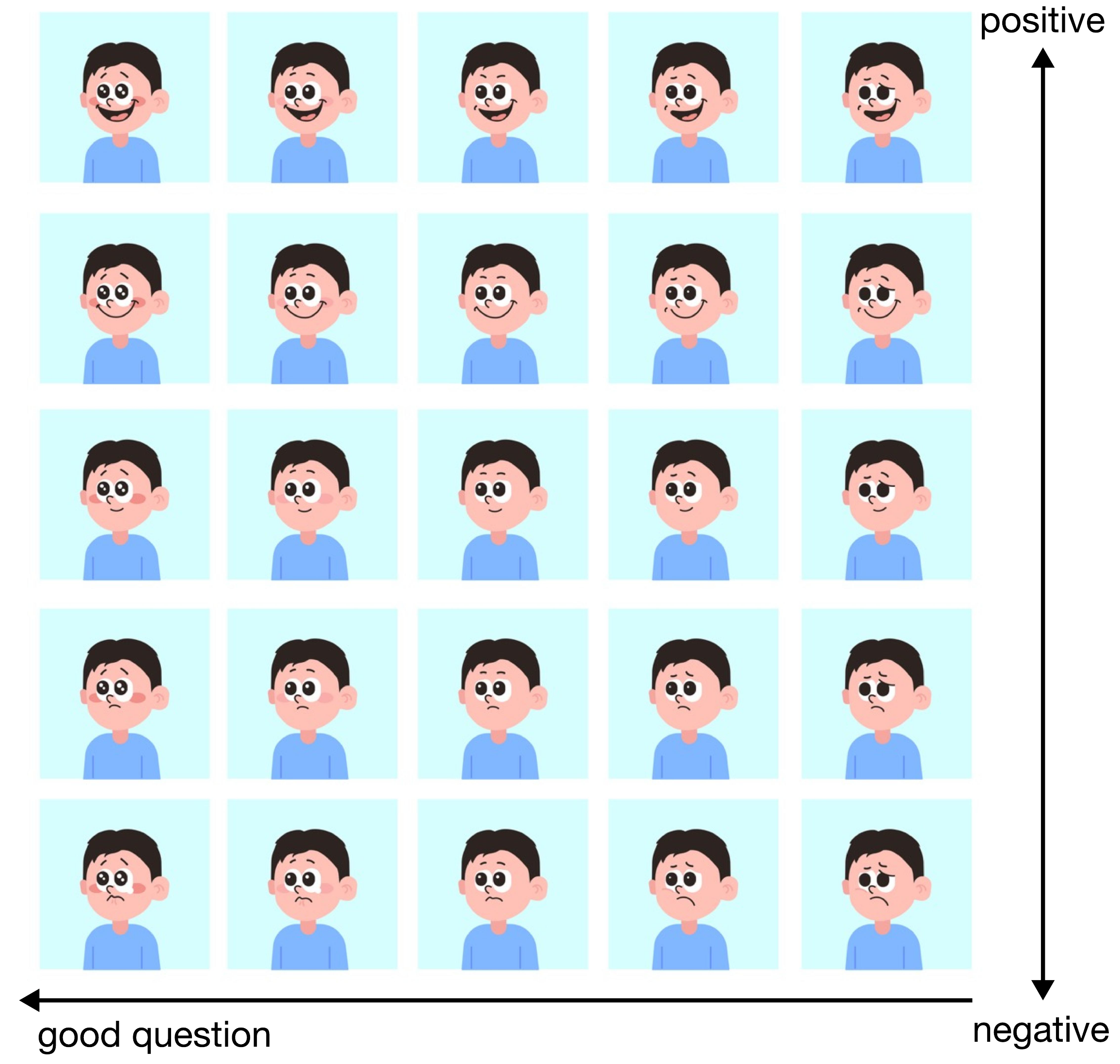}
    \caption[A grid of Alex’s faces illustrating varying emotions based on two axes: sentiment (vertical) and quality of questions (horizontal). Along the vertical axis, Alex’s facial expressions shift from negative at the bottom to positive at the top. In the lower rows, Alex’s expressions are visibly upset, with furrowed brows, frowns, and downturned eyes, indicating negative sentiment. Moving upward, the expressions become more neutral, with relaxed eyes and a slight smile, eventually turning into broad smiles with raised eyebrows at the top, representing positive sentiment. Along the horizontal axis, the quality of questions moves from poor (left) to good (right). On the far left, Alex’s expressions reflect confusion or concern, with wide eyes and tense facial muscles. As you move toward the right, his expressions grow more confident and satisfied.]{A grid of \mentee{}'s faces illustrating varying emotions based on two axes: sentiment (vertical) and quality of questions (horizontal). The facial expression of \mentee{} initially starts at the coordinate (3,3) and moves up or down by one space per feedback evaluation result. If it is already at its happiest expression, it will not change upon receiving positive feedback. The same applies in the other direction.}
    \label{fig:facial_expression}
\end{center}
\end{figure}

\clearpage

\section{Raw Usage Log}

\begin{figure} [ht]
\begin{center}
    \includegraphics[width=\textwidth]{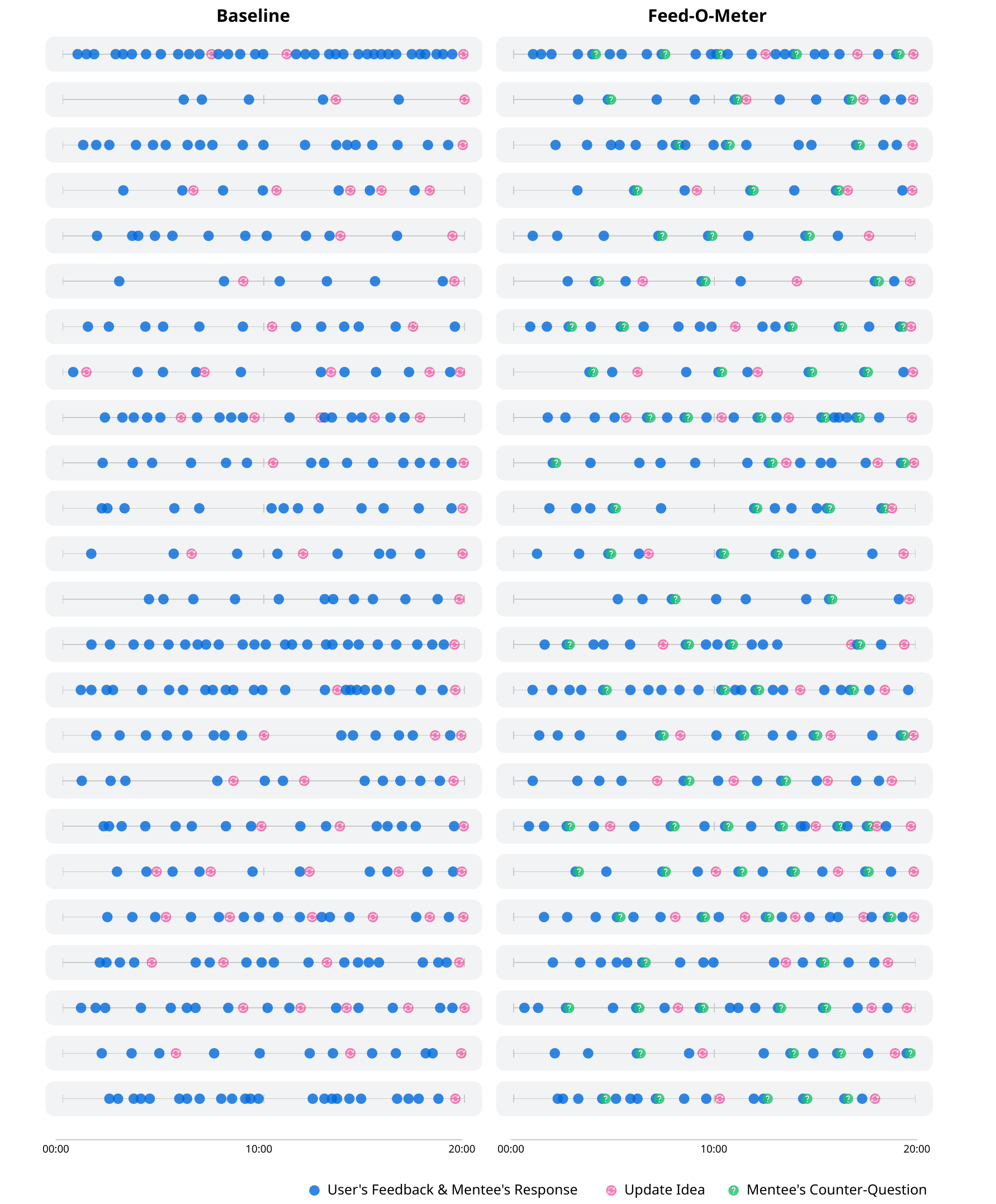}
    \caption[The figure compares the interaction timelines for 24 participants (P1 to P24) under two conditions: Baseline (left) and Feed-O-Meter (right). Each row represents a participant's timeline during a feedback session, with time indicated from 0 to 20 minutes on the horizontal axis. In both conditions, blue dots represent user feedback and mentee responses. Pink dots indicate when the user clicked the "Update Idea" button, and green dots show instances where the mentee asked a counter-question, which only appears in the Feed-O-Meter condition.]{Visualization of the usage logs from both baseline and \sysname{} conditions for each participant along a timeline. Blue dots represent feedback exchanges between the participant and mentee, pink dots indicate idea updates, and green dots (only in the \sysname{} condition) show when the mentee asked a counter-question.}
    \label{fig:feedback_logdata_by_timeline}
    \vspace{-0.5cm}
\end{center}
\end{figure}

\clearpage

\section{Final Ideas generated by AI Mentee from the User Study}

\begin{table}[h]
\scriptsize
\centering
\begin{tabular}{p{0.14\textwidth}p{0.78\textwidth}}
\toprule
\multicolumn{2}{c}{\textbf{Child Protection}} \\
\midrule
Initial Idea & \textbf{Title:} Wearable Device for Child Safety

\textbf{Problem:} Child protection services exist to prevent child abuse and neglect, and to support child safety. However, child abuse and neglect mainly occur indoors, making it difficult for bystanders to recognize and respond to a child's condition. In other words, child protection services have spatial limitations. To quickly recognize and respond to situations of child abuse or neglect, it is necessary to collect data that can identify a child's condition anywhere.

\textbf{Idea:} A child safety wearable device that looks like a regular bracelet has GPS tracking functionality, a microphone, heart rate monitoring, and an emergency call button. Real-time data about the child's situation can be collected through GPS, a microphone, and a heart rate. Therefore, if a child is in a dangerous situation, their condition can be immediately checked, and appropriate action can be taken. Additionally, if the child is in a situation where they can report themselves, they can directly press the emergency call button to request help. \\
\midrule
P18 (baseline) & \textbf{Title:} Wearable Device for Continuous Monitoring of Child's Physical Condition and Location

\textbf{Target Problem:} Child abuse mainly occurs indoors, which is an environment difficult to intervene from the outside. Also, while quick response is important in cases of child abuse, current child protection systems often fail to provide sufficient support for this. Accordingly, a practical solution is needed to detect abuse situations in a timely manner and respond quickly.

\textbf{Idea:} This wearable device is available in various forms that can be attached to a child's wrist, neck, or shoes and monitors the child's heart rate, location, and surrounding sounds in real-time. In addition to a function that notifies emergency situations with a simple button operation, it implements a function that automatically detects danger and immediately sends notifications to parents or guardians. It also provides selectable alarm methods depending on the situation, including functions that allow children to directly request help through vibration or sound. This device is designed for children to respond immediately in emergency situations and has a simple and intuitive interface necessary for detecting dangers and taking action. In the future, this device will be continuously improved based on user feedback and plans to introduce improved functions through analysis of various child abuse situations. \\
\midrule
P12 (\sysname{}) & \textbf{Title:} Customized Wearable Device for Child Safety

\textbf{Target Problem:} Wearable devices currently on the market are mainly designed for adult users, and as a result, they are not designed with sufficient consideration for children's cognitive and usage abilities. This creates limitations in children being protected in real-time using the device and quickly responding to emergency situations. Also, there are limited ways for children to quickly alert parents or guardians in potentially dangerous situations.

\textbf{Idea:} This wearable device is designed to match children's cognitive and motor skills, providing a simple and clear user interface (UI). Key features include GPS tracking, real-time microphone monitoring, heart rate measurement, and a physical emergency call button. The device's GPS tracking function is always active and easily displayed on the screen, allowing real-time location checking of the child. The emergency call button is designed to be easily accessible to children, but is protected in a special way to prevent malfunction. This device is designed considering children's characteristics, using characters and colors that children like, and provides various customization options such as allowing users to add or decorate accessories themselves. In particular, the UI uses clear color coding for functions that need to be emphasized, making it intuitively understandable. This wearable device is designed to help children maintain safety both indoors and outdoors and to respond quickly in dangerous situations. \\
\bottomrule
\end{tabular}
\caption[The table shows two examples of how participants expanded upon the same initial design idea related to child protection wearable devices. Each row contains a Title, Target Problem, and Idea. The initial design idea is titled "Wearable Device for Child Safety." Participant 18, in the baseline condition, developed the idea titled "Wearable Device for Continuous Monitoring of Child's Physical Condition and Location." Participant 12, in the Feed-O-Meter condition, developed the idea titled "Customized Wearable Device for Child Safety."]{Examples of design ideas from the user study on the topic of Child Protection.}
\label{tab:child-safety-wearable-ideas}
\end{table}

\end{document}